\documentclass[superscriptaddress,aps,pra,twocolumn,showpacs,nofootinbib]{revtex4-1}
\usepackage{etex}
\usepackage{amsmath,amssymb,amsthm}
\usepackage[colorlinks=true,citecolor=blue,urlcolor=blue]{hyperref}
\usepackage[pdftex]{graphicx}
\usepackage{times,txfonts}
\usepackage{braket}
\usepackage{color}
\usepackage{natbib}

\newcommand{\be}{\begin{equation}}
\newcommand{\ee}{\end{equation}}
\newcommand{\ba}{\begin{eqnarray}}
\newcommand{\ea}{\end{eqnarray}}
\newcommand{\ketbra}[2]{|#1\rangle \langle #2|}

\newcommand{\one}{\bf{1}}

\newcommand{\etal}{{\it{et. al. }}}

\newtheorem{result}{Result}

\newtheorem{definition}{Definition}
\newtheorem{proposition}{Proposition}

\newtheorem{Lemma}{Lemma}

\newtheorem{remark}{Remark}
\begin{document}
\title{Study of nonclassicality in tripartite correlations beyond standard Bell nonlocality paradigm}   
   \author{C.  Jebaratnam}  
\email{jebarathinam@gmail.com} 
\affiliation{S. N. Bose National Centre for Basic Sciences, Salt Lake, Kolkata 700 098, India} 
\date{\today}
\begin{abstract}
In the nonsignaling framework, nonclassicality in correlation arising from two spatially separated input-output devices gets manifested, solely, through its \emph{nonlocal} behavior. Study of correlations based on this said feature is commonly known as local-nonlocal paradigm. While in two-party scenario correlations can be of only two types either local or nonlocal, the situation gets more involved for multi-party scenario, \emph{e.g.,} for tripartite scenario, correlations can be of three types: fully local, two-way local, and genuinely nonlocal. Nonsignaling correlations having quantum realization are termed physical. Fully local and certain quantum realizable two-way local tripartite correlations always have a quantum realization with tripartite biseparable states if there is no restriction on the local Hilbert-space dimensions. In this work, we study the quantum simulation of fully local and two-way local  tripartite correlations with restricted local Hilbert-space dimensions, in particular we consider $\mathbb{C}^2\otimes\mathbb{C}^2\otimes\mathbb{C}^2$ quantum systems. Interestingly, in this restricted simulation scenario we find that simulation of certain fully local and two-way local correlations necessarily requires \emph{genuine quantumness} in the three qubit states. This, going beyond the standard nonlocality paradigm, captures a new notion of genuine nonclassicality even in the fully local and two-way local correlations. To explore this newly introduced notion of genuine nonclassicality, we propose two quantities of interest, called \emph{Svetlichny strength} and \emph{Mermin strength}, and extensively study their properties.

\end{abstract}
\pacs{03.65.Ud, 03.67.Mn, 03.65.Ta}
\maketitle
\section{Introduction}

Correlations in the outcomes of local measurements performed on entangled states are in general incompatible
with local hidden variable (LHV) models. This feature of quantum correlations is commonly known as \emph{nonlocality}
and was first pointed out by J. S. Bell in his seminal 1964 paper \cite{bell64}. Bell studied such nonlocal correlations 
in the case of bipartite scenario, i.e., for correlations arising from local measurements performed on two spatially separated systems. In the multipartite scenario (involving more than two spatially separated systems), the nonlocal behavior can be manifested in different ways \cite{BNL}. In particular, for tripartite scenario, Svetlichny showed that quantum correlations can have genuine nonlocality in the sense that they cannot be explained by a
hybrid local-nonlocal hidden variable model, and such genuine nonlocality can be witnessed by the violation of a Svetlichny inequality (SI) \cite{SI}.

It is important to note that though quantum theory (QT) exhibits nonlocal behavior, it does not contradict relativistic causality principle (also known as non-signaling (NS) principle). Moreover, QT is not the only NS theory that exhibits nonlocalily. In a seminal paper  \cite{PR}, Popescu and Rohrlich (PR) provided an example of a hypothetical correlation that is more nonlocal than QT, but nevertheless, satisfies the NS principle. This initiated the study of nonlocal correlations in a more general framework than QT known as generalized nonsignaling theories (GNST) \cite{MAG06,RevSP}. Within the framework of GNST, correlations are treated as arising from ``black boxes'' which are input-output devices. A correlation will be called quantum if it can be realized by a set of local measurements on a composite quantum system.

A bipartite quantum state must be entangled one in order to give rise to nonlocal correlations, i.e., within QT entanglement is necessary for exhibiting nonlocality. On the other hand, any local correlation always has a quantum realization with some separable quantum state. In recent past it was shown that even mixed separable states can have nonclassical correlations \cite{WZQD,WZQD1,Modietal} which motivated the study of quantum correlations beyond entanglement. Such correlations are named as quantum discord which has been shown to be useful resource for different information theoretic tasks \cite{QSM,QDRS}. Quantum states with zero discord are called classically correlated states which can always be written as a convex mixture of some orthonormal product basis states \cite{CQstate,Dakicetal}. Very recently genuine tripartite quantum discord, a generalized measure of genuine tripartite quantum correlations, has been studied in \cite{GTC,GCTS,GTQD}.

As mentioned earlier, in the tripartite scenario, violation of a SI is sufficient to certify genuine nonlocality in a correlation. 
A correlation can be non-genuine in two ways: either it is \emph{fully local} or it is \emph{two-way local} \cite{Barrett}. In this tripartite scenario, there are two-way local correlations that violate a Mermin inequality (MI) \cite{mermin}.
Note that not all two-way local correlations are quantum realizable, whereas all the fully local correlations are quantum \cite{Barrett}. Here our main interest is to study quantum realizable tripartite correlations. 
The observation of genuine nonlocality implies the presence of genuine entanglement 
in a device-independent (DI) way (i.e., without
any assumption on the measured system as well as the measurements performed).
It is known that the observation of a two-way local  correlation that violates a MI
more than $2\sqrt{2}$ also implies the presence of genuine entanglement in a DI way \cite{SU01,CGP+02,DIMulti}.
Note that all the fully local correlations and  
two-way local correlations that do not certify genuine entanglement in a DI way
can be simulated by fully separable and biseparable states respectively \cite{BNL,suplo}.
 
Nonlocal cost is a measure of nonlocality which is based on the Elitzur-Popescu-Rohrlich (EPR2) decomposition \cite{EPR2}.
In the EPR2 approach, a given 
correlation (box) is decomposed in a convex mixture of a nonlocal part and a local part, with the weight
of the nonlocal part minimized. 
For a correlation that violates a SI,
the nonlocal cost is obtained by minimizing the weight of a pure genuinely nonlocal box known as Svetlichny box. 
In the case of a (quantum realizable) two-way local correlation that violates 
a MI more than $2\sqrt{2}$, the nonlocal cost is obtained
by minimizing the weight of a pure nonlocal box which we call Mermin box. 
In this paper, we call the maximal Svetlichny-box weight of a correlation that 
violates a SI  \emph{Svetlichny strength} (SS) and the maximal Mermin-box weight of a correlation 
that violates a MI \emph{Mermin strength} (MS).  

The observation that certain fully local correlations 
have nonzero SS or nonzero MS motivates us to ask whether 
the quantum simulation of these correlations requires \emph{genuine quantumness} in the states.
The answer is no when there is no restriction on the local Hilbert-space dimensions. 
Therefore, we put the constraint that for simulating those correlations one cannot use quantum states of arbitrary 
local Hilbert-space dimensions. More specifically to say we consider only three-qubit quantum systems (i.e., quantum systems with Hilbert space $\mathbb{C}^2\otimes\mathbb{C}^2\otimes\mathbb{C}^2$) to simulate the fully local and two-way local tripartite correlations. In this restricted simulation scenario, we find the following interesting observations: 
\begin{itemize}
\item[(i)]  certain two-way local correlations 
cannot be simulated by biseparable states of $\mathbb{C}^2\otimes\mathbb{C}^2\otimes\mathbb{C}^2$ systems
which do not have genuine quantum discord;
\item[(ii)] moreover, there exist two-way local correlations that cannot be simulated with quantum states of $\mathbb{C}^2\otimes\mathbb{C}^2\otimes\mathbb{C}^2$ systems having genuine quantum discord, rather genuine entanglement is necessary, even if they do not certify genuine entanglement in a DI way;
\item[(iii)] even for simulation of certain fully local correlations  by $\mathbb{C}^2\otimes\mathbb{C}^2\otimes\mathbb{C}^2$ systems, one requires genuine quantum discord states.
\end{itemize}
These observations motivate us to characterize the (genuine) nonclassicality present in the local as well as two-way local correlations by using the two quantities SS and MS. Our study thus goes beyond the standard locality-nolocality paradigm, since it captures the (genuine) nonclassicality present in the fully local as well as two-way local correlations: the quantum states must contain genuine quantum correlations in the form of genuine discord, when those correlations are supposed to be simulated by quantum states of restricted local dimensions. For a general tripartite NS correlation  belonging to \emph{Svetlichny polytope}, a proper sub-polytope of the full NS polytope, we obtain a canonical decomposition with three different boxes: (a) a Svetlichny box which  violates the SI maximally, (b) a two-way local box that violates the MI maximally, and (c) a box that belongs to a subregion of the two-way local polytope. In this decomposition, the weights of the boxes (a) and (b) are called SS and MS respectively.
In the canonical decomposition, the presence of genuine nonclassicality is manifested in three different ways: when only SS is nonzero, or only MS is nonzero, or both SS and MS are nonzero.

The paper is organized as follows. In Sec. \ref{prl}, we review the geometry of the full tripartite nonsignaling polytope.
In Sec. \ref{Spoly}, we restrict ourselves to the Svetlichny-box polytope and we introduce two notions of genuine nonclassicality, namely \emph{Svetlichny strength} and \emph{Mermin strength}. 
In Sec. \ref{QC},  we analyze genuine nonclassicality of boxes arising from three-qubit states with respect to these two notions. Lastly, we present our conclusions in Sec. \ref{conc}.
\section{Preliminaries}\label{prl}
We consider the following tripartite scenario. Three spatially separated parties, say Alice, Bob, and Charlie, share a tripartite NS box with two binary inputs and two binary outputs per party. Let $x$, $y$, and $z$ denote the inputs and $a$, $b$, and $c$ denote the outputs of Alice, Bob, and Charlie respectively, where $x,y,z,a,b,c\in \{0,1\}$. The correlation between the outputs is captured by the set of conditional probability distributions (CPDs), $P(abc|xyz)$.
In addition to positivity and normalization, the CPDs satisfy NS constraints, that read,
\be
\sum_a P(abc|xyz)=P(bc|yz) \quad \forall b,c,x,y,z,
\ee
and permutations of the parties, i.e., marginal distribution of any two parties of the box is independent of the input chosen by the other party.
The set of such NS boxes forms a convex polytope $\mathcal{N}$ in a $26$ dimensional space \cite{Barrett}. Any box belonging to this polytope can be fully specified by $6$ single-party, $12$ two-party and $8$ three-party expectations,
\begin{align}
&P(abc|xyz)=\frac{1}{8}[1+(-1)^a\braket{A_x}+(-1)^b\braket{B_y}+(-1)^c\braket{C_z}\nonumber \\
&+(-1)^{a\oplus b}\braket{A_xB_y}+(-1)^{a\oplus c}\braket{A_xC_z}+(-1)^{b\oplus c}\braket{B_yC_z}\nonumber \\
&+(-1)^{a\oplus b\oplus c}\braket{A_xB_yC_z}],
\end{align}
where $\braket{A_x}=\sum_a (-1)^a P(a|x)$, $\braket{A_xB_y}=\sum_{a,b}(-1)^{a\oplus b}P(ab|xy)$ and the other expectations are similarly defined, $\oplus$ denotes modulo sum $2$.

The set of boxes that can be simulated by a fully LHV model are of the form,
\begin{align}
P(abc|xyz)=\sum_\lambda p_\lambda P_\lambda(a|x)P_\lambda(b|y)P_\lambda(c|z), \label{LHV}
\end{align}
and they form a local polytope \cite{LHV,WernerWolfmulti} denoted $\mathcal{L}$. Here, $\lambda$ denotes shared classical randomness which occurs with probability $p_\lambda$.
The extremal boxes of $\mathcal{L}$ are $64$ local vertices which are fully deterministic boxes,
\be
P^{\alpha\beta\gamma\epsilon\zeta\eta}_D(abc|xyz)=\left\{
\begin{array}{lr}
1, & a=\alpha x\oplus \beta\\
   & b=\gamma y\oplus \epsilon \\
   & c=\zeta  z\oplus \eta\\
0 , & \text{otherwise}.\\
\end{array}
\right.  \label{DB} \ee
Here, $\alpha,\beta,\gamma, \epsilon, \zeta, \eta \in\{0,1\}$. These deterministic boxes can be written as the product of marginals corresponding to each party, i.e.,
$P_D(abc|xyz)=P_D(a|x)P_D(b|y)P_D(c|z)$.
Boxes lying outside $\mathcal{L}$ are called nonlocal boxes and they cannot be written as a convex mixture of the local deterministic boxes alone.

In Ref. \cite{Pironioetal}, Pironio \etal found that the full NS polytope $\mathcal{N}$ has $53856$ extremal boxes (vertices) which can be classified
into 45 different classes of nonlocal vertices and a class comprising local vertices.
The vertices in each class are equivalent
in the sense that they can be converted into each other through local reversible operations (LRO)
and permutations of the parties. LRO, which are analogous to local unitary operations in quantum theory,
include local relabeling of the inputs and outputs.
For instance, the operations $x\rightarrow x \oplus 1$
and $a \rightarrow a \oplus \alpha x \oplus \beta$  on Alice's side are local reversible.

Nonlocal boxes can be classified into genuinely three-way nonlocal and two-way local boxes.
A nonlocal box is genuinely three-way nonlocal \emph{if and only if} (iff) it cannot be simulated by a
hybrid local-nonlocal \textit{NS}  model \cite{Banceletal},
\begin{align}
P(abc|xyz)&=p_1\sum_\lambda p_\lambda P_\lambda^{AB|C}+p_2\sum_\lambda q_\lambda P_\lambda^{AC|B} \nonumber \\
&+p_3\sum_\lambda r_\lambda P_\lambda^{A|BC}, \label{HLNL}
\end{align}
where $P_\lambda^{AB|C}=P_\lambda(ab|xy)\,P_\lambda(c|z)$, and, where $P_\lambda^{AC|B}$ and
$P_\lambda^{A|BC}$  are  similarly  defined.
Each bipartite distribution in this decomposition can have arbitrary nonlocality consistent with the NS principle.

The set of boxes that admit a decomposition as in Eq. (\ref{HLNL}) again forms a convex polytope which is called two-way local polytope denoted $\mathcal{L}_2$.
The extremal boxes of this polytope are the $64$ local vertices and $48$ two-way local vertices which are the bipartite PR-boxes.
There are $16$ two-way local vertices in which a PR-box \cite{PR} is shared between $A$ and $B$,
\begin{align}
&P^{\alpha\beta\gamma\epsilon}_{12}(abc|xyz) \nonumber \\
&=\left\{
\begin{array}{lr}
\frac{1}{2}, & a\oplus b=x\cdot y \oplus \alpha x\oplus \beta y \oplus \gamma \quad \& \quad o=\gamma k \oplus \epsilon\\
0 , & \text{otherwise},\\
\end{array}
\right. \label{PR}
\end{align}
and the other $32$ two-way local vertices, $P^{\alpha\beta\gamma\epsilon}_{13}$ and $P^{\alpha\beta\gamma\epsilon}_{23}$,
in which a PR-box is shared by $AC$ and $BC$ are similarly defined. The extremal boxes in Eq. (\ref{PR}) can be written in the factorized form,
$P^{\alpha\beta\gamma\epsilon}_{12}(abc|xyz)=P^{\alpha\beta\gamma}_{PR}(ab|xy)P^{\gamma\epsilon}_D(c|z)$, where
$P^{\alpha\beta\gamma}_{PR}(ab|xy)$ are the $8$ PR-boxes given by the relation in Eq. (\ref{PR})
and $P^{\gamma\epsilon}_D(c|z)=\delta^z_{c\oplus \gamma z \oplus \epsilon}$.
The set of two-way local boxes satisfy, $\mathcal{L} \subset \mathcal{L}_2 \subset \mathcal{N}$.
A three-way nonlocal box cannot be written as a convex mixture of the extremal boxes of $\mathcal{L}_{2}$ alone and violates
a facet inequality of $\mathcal{L}_2$.
Bancal \etal \cite{Banceletal}
found that $\mathcal{L}_2$ has $185$ classes of facet inequalities which include Bell-type inequalities detecting three-way nonlocality.

In this work, we consider two classes of $3$-way nonlocal vertices that belong to the classes $8$ and $46$ given in the table of Pironio \etal \cite{Pironioetal}.
The extremal three-way nonlocal boxes that belong to
the class $46$ are $16$ Svetlichny-boxes which are given as follows:
\begin{align}
&P^{\alpha\beta\gamma\epsilon}_{\rm Sv}(abc|xyz)  \nonumber \\
&=\left\{
\begin{array}{lr}
\frac{1}{4}, & \!a\!\oplus \!b\!\oplus \!c\!
=\!x\cdot y \!\oplus \!x\cdot z\! \oplus \!y\cdot z \!\oplus \!\alpha x\!\oplus\! \beta y\! \oplus\! \gamma z \!\oplus\! \epsilon\\
0 , & \text{otherwise}.\\
\end{array}
\right. \label{NLV}
\end{align}
A Svetlichny box, $P^{\alpha\beta\gamma\epsilon}_{\rm Sv}$, violates one of the class $185$ facet inequalities,
\be
\mathcal{S}_{\alpha\beta\gamma\epsilon}
=\sum_{xyz}(-1)^{x\cdot y \oplus x\cdot z \oplus y\cdot z \oplus \alpha x\oplus \beta y \oplus \gamma z \oplus \epsilon}\braket{A_xB_yC_z}\le4, \label{SI}
\ee
to its algebraic maximum of $8$. A class $185$ facet inequality is known as Svetlichny inequality  which detects the strongest form of three-way nonlocality \cite{SI}.
\begin{definition}
A NS box is said to exhibit Svetlichny nonlocality if it violates a Svetlichny inequality.
\end{definition}
We will refer to the boxes which do not violate a SI as \emph{Svetlichny-local}.

A tripartite NS box is quantum if it can be expressed through the Born's rule, i.e.,
\be
P(abc|xyz)=\mathrm{Tr} \left(\rho M^{a}_{A_x}\otimes M^{b}_{B_y}\otimes M^{c}_{C_z}\right),
\ee
where $\rho$ is a tripartite quantum state (density matrix) acting on the composite Hilbert space $\mathcal{H}_A\otimes\mathcal{H}_B\otimes\mathcal{H}_C$, where $\mathcal{H}_K$ denotes Hilbert space of
$k$th party, and $M^{a}_{A_x}$,  $M^{b}_{B_y}$, and $M^{c}_{C_z}$ are the local measurement operators which are in general  positive operator valued measurements (POVMs). Here dimension of each local Hilbert space $\mathcal{H}_K$ is arbitrary. In the following we consider simulation of boxes
$P:= P(abc|xyz) \in \mathcal{L}_2$ with $\mathbb{C}^2\otimes\mathbb{C}^2\otimes\mathbb{C}^2$ quantum states,
i.e., three-qubit states. Interestingly, we find that there exist $P\in \mathcal{L}_2$ which allow a $\mathbb{C}^2\otimes\mathbb{C}^2\otimes\mathbb{C}^2$ quantum simulation only if the quantum states possess some sort of genuine quantum correlations measured in terms of genuine quantum discord \cite{GTC,GCTS,GTQD}. And in this sense we say that the studied correlations contain genuine nonclassicality (note that this goes beyond the conventional nonlocality paradigm).

\section{Svetlichny strength and Mermin strength} \label{Spoly}
In this section, we introduce the quantities Svetlichny strength (SS) and Mermin strength (MS), which can be nonzero even for boxes admitting LHV model. In the case of local boxes arising from three-qubit states (i.e., boxes which can be simulated by  $\mathbb{C}^2\otimes\mathbb{C}^2\otimes\mathbb{C}^2$ quantum systems),
nonzero SS and nonzero MS characterize genuine nonclassicality. Correlations having nonzero SS we call Svetlichny nonclassical and similarly Mermin nonclassical whenever MS is nonzero. For the purpose of our study, we restrict ourselves in a subpolytope called Svetlichny-box polytope which we describe below.
\begin{figure}
\centering
\includegraphics[width=0.45\textwidth]{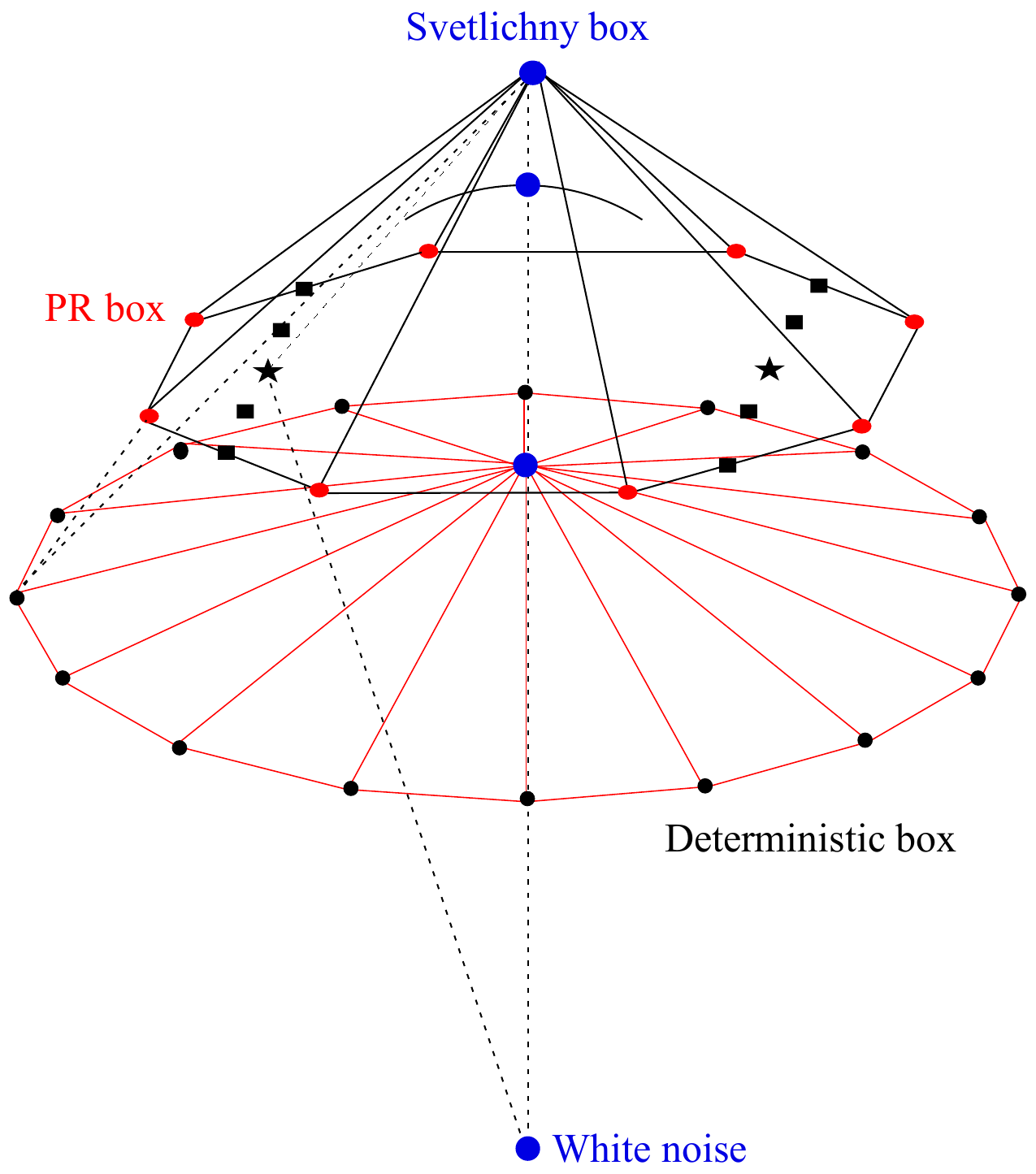}
\caption{A three-dimensional representation of the Svetlichny-box polytope is shown here. The fully deterministic boxes are represented by the circular points on the hexadecagon. The bipartite PR-boxes are represented by the circular points on the octagon. The circular point on the top represents the Svetlichny box. The region that lies above the hexadecagon and below the octagon represents the two-way local region. The region below the curved surface contains quantum boxes and the point on this curved surface represents the quantum box that achieves maximal Svetlichny nonlocality. The star and square points represent Mermin boxes with maximally and nonmaximally mixed bipartite marginals, respectively. The triangular region (shown by dotted lines) which is a convex hull of the Svetlichny box, the Mermin box and white noise illustrates Result $2$ in the main text: any point that lies inside the triangle can be decomposed into Svetlichny box, the Mermin box and white noise. The circular point at the center of the hexadecagon is the isotropic Svetlichny-box with $p_{Sv}=\frac{1}{2}$ which can be decomposed as an equal mixture of the $16$ deterministic boxes or an equal mixture of two Mermin boxes. Note that this figure is illustrative and inspired by the figure depicting the PR-box polytope given in Ref. \cite{DQKD}; the actual polytope is $26$-dimensional. }\label{NS3dfig}
\end{figure}
\subsection{Svetlichny-box polytope}
The Svetlichny-box polytope denoted $\mathcal{R}$ has $128$ extremal boxes which are the Svetlichny boxes, the bipartite PR-boxes and the deterministic boxes. Any box $P:=P(abc|xyz)\in\mathcal{R}$ can be written as a convex mixture of these extremal boxes, i.e.,
\begin{equation}
P=\sum^{15}_{i=0}p_iP^i_{Sv}+\sum^{15}_{i=0}q_iP^i_{12}+\sum^{15}_{i=0}r_iP^i_{13}+\sum^{15}_{i=0}s_iP^i_{23}+\sum^{63}_{j=0}t_jP^j_{D},
\label{eq:gendecomp}
\end{equation}
with $\sum_ip_i+\sum_iq_i+\sum_ir_i+\sum_is_i+\sum_jt_j=1$. Here, $P^j_{D}$, $P^i_{12}$, and $P^i_{Sv}$  are 
the boxes given in Eqs. (\ref{DB}), (\ref{PR}), and (\ref{NLV}), respectively, with $i=2^3\alpha+2^2\beta+2^1\gamma+2^0\epsilon$ and $j=2^5\alpha+2^4\beta+2^3\gamma+2^2\epsilon+2^1\zeta+2^0\eta$.
The Svetlichny-box polytope can be divided into a Svetlichny nonlocal region and the two-way local polytope.
We have that $\mathcal{L} \subset \mathcal{L}_2 \subset \mathcal{R} \subset \mathcal{N}$. Diagrammatically the Svetlichny-box polytope is shown in Fig. \ref{NS3dfig}.

\subsection{Svetlichny strength}
We consider isotropic Svetlichny-box defined as
\be
P=\frac{2+abc(-1)^{xy\oplus xz \oplus yz \oplus x \oplus y \oplus z}\mu}{16}; \quad 0<\mu\le 1. \label{SvF}
\ee
The above box violates the Svetlichny inequality (given in Eq. (\ref{SI})) 
for $\mu>\frac{1}{2}$, and for $\mu\le\frac{1}{2}$ it is fully local as in this range the correlation does not violate any Bell inequality \footnote{A tripartite correlation $P\in\mathcal{L}_2$ is nonlocal \emph{iff} it violates either a MI or a Bell--Clauser-Horne-Shimony-Holt (CHSH) inequality \cite{chsh,WernerWolf}.}. This implies that for $\mu\le1/2$, the isotropic Svetlichny-box can be
decomposed in a convex mixture of the local deterministic boxes alone. However, the box (\ref{SvF}) can be 
expressed as the mixture
\be
P=p_{Sv}P^{0000}_{Sv}+(1-p_{Sv})P_N; \quad 0<p_{Sv}\le 1, \label{nSv}
\ee
with $p_{Sv}=\mu$. Here,  $P_N$ is the maximally mixed box, i.e., $P_N(abc|xyz)=1/8$ for all $x,y,z,a,b,c$.
We can see that even when the isotropic Svetlichny-box is local, it has nonvanishing Svetlichny-box weight $p_{Sv}$
originating from Svetlichny nonlocality for any $\mu>0$ (as it lies on the line segment joining the Svetlichny-box and 
white noise). 
We call the maximal such irreducible weight of the Svetlichny-box, which occurs above a Svetlichny-local box 
even in the case of two-way local and fully local boxes, \textit{Svetlichny strength}.

Consider the modulus of the Svetlichny functions in Eq. (\ref{SI}), i.e., $\mathcal{S}_{\alpha\beta\gamma}:=|\mathcal{S}_{\alpha\beta\gamma\epsilon}|$.
For any isotropic Svetlichny-box,
\be
P=p_{Sv}P^{\alpha\beta\gamma\epsilon}_{Sv}+(1-p_{Sv})P_N; \quad 0<p_{Sv}\le 1, \label{nSv}
\ee
it can be checked that one of the Svetlichny functions  $\mathcal{S}_{\alpha\beta\gamma}(P)=8p_{Sv}$
and the others take the value zero. Any such isotropic Svetlichny-box is an unequal
mixture of the Svetlichny-box $P^{\alpha\beta\gamma\epsilon}_{Sv}$ and the anti-Svetlichny-box  
$P^{\alpha\beta\gamma\epsilon\oplus1}_{Sv}$ whose uniform mixture gives white noise, i.e., the box
can also be decomposed as $P=\frac{1+p_{Sv}}{2}P^{\alpha\beta\gamma\epsilon}_{Sv}+\frac{1-p_{Sv}}{2}P^{\alpha\beta\gamma\epsilon\oplus1}_{Sv}$.
Thus, we see that nonzero value of the quantity  $\mathcal{S}_{\alpha\beta\gamma}(P)$ 
measures the Svetlichny strength of any isotropic Svetlichny-box (\ref{nSv}).

By using the Svetlichny functions $\mathcal{S}_{\alpha\beta\gamma}$, 
we now introduce a quantity denoted $\mathcal{G}$ which is defined as
\be
\mathcal{G}:=\min\{\mathcal{G}_1,...,\mathcal{G}_9\}, \label{GBD}
\ee
where $\mathcal{G}_1:=\Big|\Big||\mathcal{S}_{000}-\mathcal{S}_{001}|-|\mathcal{S}_{010}-\mathcal{S}_{011}|\Big|
-\Big||\mathcal{S}_{100}-\mathcal{S}_{101}|-|\mathcal{S}_{110}-\mathcal{S}_{111}|\Big|\Big|$ and the other eight $\mathcal{G}_i$ are similarly defined and can be obtained by interchanging $\mathcal{S}_{\alpha\beta\gamma}$'s in $\mathcal{G}_1$. The quantity $\mathcal{G}$ satisfies the following properties: (g1) positivity, i.e., $\mathcal{G}\ge0$, (g2) the bipartite PR-boxes and the deterministic boxes have $\mathcal{G}=0$, (g3) $\mathcal{G}$ is invariant under LRO since the set $\{\mathcal{G}_i\}$ is invariant under LRO, and (g4) the algebraic maximum of $\mathcal{G}$ is achieved by the Svetlichny boxes, i.e., $\mathcal{G}=8$ for any Svetlichny box.

By using the quantity $\mathcal{G}$, we will now derive a canonical decomposition for any box $P\in\mathcal{R}$.       
Let   a   decomposition   (\ref{eq:gendecomp})   for
$P$    be   
\be
P    =    \sum^{15}_{i=0} g_i P^i_{Sv}  +
\left(1-\sum^{15}_{i=0}g_i\right)P_{ SvL}. \label{st1}
\ee
In this decomposition, the Svetlichny-box components are maximized 
over all possible decompositions such that the remaining Svetlichny-local
box does not have any Svetlichny-box component.    
We now express the first term in  the decomposition (\ref{st1}) as a sum of the Svetlichny-local boxes, $P^i_{SvL}$,
and a \emph{residual} box, $\textbf{P}^\uparrow_{\rm Sv}$ (the dominant Svetlichny-box), i.e.,
\be
\sum^{15}_{i=0} g_i P^i_{Sv}=\mu \textbf{P}^\uparrow_{\rm Sv}+\sum^{15}_{i=1} p_iP^i_{SvL}. \label{st2}
\ee
This is done by recursively rewriting the mixture of two different Svetlichny boxes, $pP^i_{Sv}+qP^j_{Sv}$ ($p>q$),
as the mixture $(p-q)P^i_{Sv}+2qP_{SvL}$, where $P_{SvL}\equiv(P^i_{Sv}+P^j_{Sv})/2$ which is Svetlichny-local. 
Since finding the above residual decomposition is not unique, $\mu$ in Eq. (\ref{st2}) can take different value for different residual decomposition. 
The evaluation of $\mathcal{G}_1$ for the mixture of the Svetlichny-boxes gives,  
\begin{align}
&\mathcal{G}_1\left(\sum^{15}_{i=0} g_i P^i_{Sv}\right)\nonumber \\
&=8\Big||\Big||g_0-g_1|-|g_2-g_3|\Big|-\Big||g_4-g_5|-|g_6-g_7|\Big||\nonumber \\
&-|\Big||g_8-g_9|-|g_{10}-g_{11}|\Big|-\Big||g_{12}-g_{13}|-|g_{14}-g_{15}|\Big||\Big|.
\end{align} 
Thus, we see that $\mathcal{G}\left(\sum^{15}_{i=0} g_i P^i_{Sv}\right)/8$ gives the (minimal) weight of the irreducible Svetlichny-box $\textbf{P}^\uparrow_{\rm Sv}$ that occurs above a Svetlichny-local box.
For this reason, among all residual decompositions as in Eq. (\ref{st2})
we take that particular decomposition where $\mu=\mathcal{G}\left(\sum^{15}_{i=0} g_i P^i_{Sv}\right)/8$. 
Plugging this minimal residual decomposition in Eq. (\ref{st1}), we get the following decomposition:
\be
P=\mu \textbf{P}^\uparrow_{\rm Sv}+(1-\mu)P'_{SvL}, \label{canonical}
\ee
where $P'_{SvL}=\frac{1}{1-\mu}\left\{\sum^{15}_{i=1} p_iP^i_{SvL}+\left(1-\sum^{15}_{i=0} g_i\right)P_{SvL}\right\}$.
The Svetlichny-local box in the decomposition (\ref{canonical}) does not
have any nonzero (irreducible) dominant Svetlichny-box  weight, 
since $\mu$ isolates the maximal irreducible Svetlichny-box weight of the correlation. 
Thus, we can state the following canonical decomposition.
\begin{result}
Any box $P$ that belongs to the Svetlichny-box polytope has a decomposition of the following form:
\be
P=\mu \textbf{P}^\uparrow_{\rm Sv} +(1-\mu)P^{\mu=0}_{SvL}, \label{canoG>}
\ee
where $\textbf{P}^\uparrow_{\rm Sv}$ is the dominant Svetlichny-box and
$P^{\mu=0}_{SvL}$ is a Svetlichny-local box which does not have any nonzero (irreducible) dominant  Svetlichny-box weight. The coefficient $\mu$ is called the Svetlichny strength of the correlation.
\end{result}

 The quantity SS is different from \emph{nonlocal cost} which is the well-known quantifier of nonlocality  \cite{EPR2,EPR2B}. Whereas in the definition of nonlocal cost, one has to minimize some particular pure nonlocal box fraction (\emph{e.g.,} for bipartite case PR-box fraction, for tripartite case Svetlichny-box fraction) over all possible decompositions, but to find SS one has to minimize Svetlichny-box fraction only over all possible residual decompositions as in Eq. (\ref{st2}). Since for SS minimization is performed over a restricted set of decompositions in comparison to the case of nonlocal cost, SS can be nonzero even when nonlocal cost (in this case Svetlichny-nonlocal cost) of a correlation is zero. For instance, 
the box (\ref{SvF}) which has Svetlichny-nonlocal cost $C_{Snl}(P)=2p_{Sv}-1$ has SS $\mu=p_{Sv}$ which is nonzero even if the box is fully local.

\begin{proposition}
If a box has $\mathcal{G}=0$, then it necessarily has SS $\mu=0$. 
\end{proposition}
\begin{proof}
The statement of this proposition follows from the fact that for any $P$ as given by the decomposition (\ref{st1}), SS $\mu=\mathcal{G}\left(\sum^{15}_{i=0} g_i P^i_{Sv}\right)/8$. 
\end{proof}

It should be noted that there can be boxes which do not have SS, but nevertheless, have $\mathcal{G}>0$.
For those boxes which lie on the line segment joining a Svetlichny-box and a Svetlichny-local box with 
$\mathcal{G}=0$, the quantities SS and $\mathcal{G}$ are related to each other as follows.     
\begin{proposition}
For any given box $P\in \mathcal{R}$ which has a decomposition of the form,
\be
P=\mu \textbf{P}^\uparrow_{\rm Sv}+(1-\mu)P^{\mathcal{G}=0}_{SvL}, \label{canoG>1}
\ee
the quantities $\mathcal{G}$ defined in Eq. (\ref{GBD}) and the SS satisfy the relationship 
$\mathcal{G}(P)=8\mu$.
\end{proposition}
\begin{proof}
The nonextremal boxes in the two-way local polytope can have
the following three types of decompositions due to the convexity of $\mathcal{R}$:
(i) a convex mixture of two $\mathcal{G}=0$ boxes,
(ii) a convex mixture of two $\mathcal{G}>0$ boxes, and
(iii) a convex mixture of a $\mathcal{G}>0$ box and a $\mathcal{G}=0$ box.
Since certain convex mixture of the $\mathcal{G}=0$ boxes ($\mathcal{G}>0$ boxes) have $\mathcal{G}>0$ ($\mathcal{G}=0$),
$\mathcal{G}$ is, in general, not linear for the two decompositions (i) and (ii). However, $\mathcal{G}$ is linear for the decomposition (iii) which
implies that the box given by the decomposition (\ref{canoG>1}) has
$\mathcal{G}(P)=\mu \mathcal{G}\left(\textbf{P}^\uparrow_{\rm Sv}\right)+(1-\mu)\mathcal{G}\left(P^{\mathcal{G}=0}_{SvL}\right)=8\mu$.
\end{proof}

\begin{proposition}
The Svetlichny strength $\mu$  of any box is invariant under LRO and permutations of the parties. 
\end{proposition}
\begin{proof}
The statement of this proposition follows from the fact that $\mathcal{G}$ is invariant under LRO
and permutations of the parties.
\end{proof}
\subsection{Mermin strength}
Likewise SS, here we define another quantity called Mermin strength. For this purpose, we require the Mermin inequalities \cite{mermin,WernerWolfmulti}. MIs are the Bell-type inequalities which can detect two-way local boxes while SIs fail to do so. For the tripartite scenario, MIs are given by, 
\be
\mathcal{M}_{\alpha\beta\gamma\epsilon}=
(\alpha\oplus\beta\oplus\gamma\oplus1)\mathcal{M}^+_{\alpha\beta\gamma\epsilon}+(\alpha\oplus\beta\oplus\gamma)\mathcal{M}^-_{\alpha\beta\gamma\epsilon}\le2, \label{MI}
\ee
where $\mathcal{M}^+_{\alpha\beta\gamma\epsilon}:=(-1)^{\gamma\oplus\epsilon}\braket{A_0B_0C_1}
+(-1)^{\beta\oplus\epsilon}\braket{A_0B_1C_0}+(-1)^{\alpha\oplus\epsilon}\braket{A_1B_0C_0}
+(-1)^{\alpha\oplus\beta\oplus\gamma\oplus\epsilon\oplus1}\braket{A_1B_1C_1}$ and $\mathcal{M}^-_{\alpha\beta\gamma\epsilon}:=(-1)^{\alpha\oplus\beta\oplus\epsilon\oplus 1}
\braket{A_1B_1C_0}+(-1)^{\alpha\oplus\gamma\oplus\epsilon\oplus 1}\braket{A_1B_0C_1}
+(-1)^{\beta\oplus\gamma\oplus\epsilon\oplus 1}\braket{A_0B_1C_1}+(-1)^{\epsilon}\braket{A_0B_0C_0}$. By using the modulus of the Mermin functions in Eq. (\ref{MI}), i.e., $\mathcal{M}_{\alpha\beta\gamma}:=|\mathcal{M}_{\alpha\beta\gamma\epsilon}|$, we introduce a quantity denoted $\mathcal{Q}$ which is defined as
\be
\mathcal{Q}:=\min\{\mathcal{Q}_1,...,\mathcal{Q}_9\}, \label{GMD}
\ee
where $\mathcal{Q}_1:=\Big|\Big||\mathcal{M}_{000}-\mathcal{M}_{001}|-|\mathcal{M}_{010}-\mathcal{M}_{011}|\Big| -\Big||\mathcal{M}_{100}-\mathcal{M}_{101}|-|\mathcal{M}_{110}-\mathcal{M}_{111}|\Big|\Big|$ 
and the other eight $\mathcal{Q}_i$ are similarly defined and can be obtained by interchanging $\mathcal{M}_{\alpha\beta\gamma}$'s in $\mathcal{Q}_1$. The quantity $\mathcal{Q}$ satisfies the following properties: (q1) positivity, i.e., $\mathcal{Q}>0$, (q2) $\mathcal{Q}=0$ for the Svetlichny-boxes, bipartite PR-boxes and deterministic boxes, (q3) $\mathcal{Q}$ is invariant under LRO since the set $\{\mathcal{Q}_i\}$ is invariant under LRO, (q4) the algebraic maximum of the quantity $\mathcal{Q}$ is $4$, and boxes achieving this algebraic maximum value we call \emph{Mermin box}. As shown in \cite{Pironioetal}, the box $P^{mm}_M$ which is the uniform mixture of the two Svetlichny boxes $P^{0000}_{Sv}$ and $P^{1110}_{Sv}$ exhibits Greenberger-Horne-Zeilinger (GHZ) paradox \cite{GHZ}. It turns out that for the box $P^{mm}_M$ the quantity $\mathcal{Q}$ takes the value $4$ (hence it is an example of Mermin box) while $\mathcal{G}=0$ for this box, here the superscript `mm' is used to denote the fact that all the bipartite marginals are maximally mixed (mm), i.e., the white noise distribution. We also note that $P^{mm}_M$ allows a two-way local decomposition given as follows:
\be
P^{mm}_{M}=\frac{1}{4}\sum^4_{\lambda=1} P_\lambda(a|x)P_\lambda(bc|yz), \label{Mbdec}
\ee
where $P_1(a|x)\!=\!\delta^x_{a\oplus x}$, $P_2(a|x)=\delta^x_{a\oplus x\oplus1}$, $P_3(a|x)=\delta^x_{a\oplus 1}$, $P_4(a|x)=\delta^x_a$,
$P_1(bc|yz)\!=\!P_{PR}^{110}(bc|yz)$, $P_2(bc|yz)=P_{PR}^{111}(bc|yz)$, $P_3(bc|yz)=P_{PR}^{001}(bc|yz)$ and 
$P_4(bc|yz)=P_{PR}^{000}(bc|yz)$. This shows that correlations exhibiting the GHZ paradox are two-way local \footnote{The other $15$ two-way local boxes that exhibit the GHZ paradox can be obtained from the box (\ref{Mbdec}) by LRO.}. However, there also exist Mermin boxes (i.e., boxes with $\mathcal{Q}=4$) which do not have maximally mixed bipartite marginals. One such example is,
\be
P^{nmm}_M=\frac{1}{2}\sum^2_{\lambda=1} P_\lambda(a|x)P_\lambda(bc|yz), \label{Mbnmm}
\ee
where
$P_1(a|x)\!=\!\delta^x_{a\oplus x}$, $P_2(a|x)=\delta^x_{a\oplus 1}$,
$P_1(bc|yz)\!=\!P_{PR}^{110}(bc|yz)$, and $P_2(bc|yz)=P_{PR}^{001}(bc|yz)$. Likewise the fact that a Svetlichny box violates only one of the SIs, a Mermin box violates only one of the MIs.

We consider isotropic Mermin-box defined as
\be
P=\frac{1+abc\delta_{x\oplus y,z}\nu}{8}; \quad 0<\mu\le 1. \label{MeF}
\ee
The above box is two-way local for $\nu>\frac{1}{2}$ as it violates the Mermin inequality (given in Eq. (\ref{MI})) 
in this range, and for $\nu\le\frac{1}{2}$ it is fully local as in this range the correlation does not violate any Bell inequality. This implies that for $\nu\le1/2$, the isotropic Mermin-box  can be
decomposed in a convex mixture of the local deterministic boxes alone. However, it can be decomposed in a convex
mixture of the Mermin box $P^{mm}_M=\frac{1}{2}\left(P^{0000}_{Sv}+P^{1110}_{Sv}\right)$ and white noise 
for any $\nu>0$ as follows:
\be
P=p_{M}P^{mm}_M+(1-p_{M})P_N; \quad 0<p_{M}\le 1, \label{nMe}
\ee
with $p_M=\nu$. We can see that even when the isotropic Mermin-box is local, it has nonvanishing Mermin-box weight $p_{M}$
originating from nonlocality for any $\mu>0$ (as it lies on the line segment joining the Mermin-box and 
white noise). 
We call the maximal such irreducible Mermin-box weight, which dominates over a Svetlichny-local/nonlocal box 
even in the case of fully local boxes,  \emph{Mermin strength}. The isotropic Mermin-box has MS $\nu=p_M$, whereas
nonlocal cost \cite{EPR2B} (in this case Mermin-nonlocal cost) $C_{Mnl}(P)=(2p_M-1)$.  

We are now in a position to obtain a more general version of Result $1$.
\begin{result}
Any box $P\in\mathcal{R}$ has a decomposition of the following form: 
\be
P=\mu \textbf{P}^\uparrow_{\rm Sv}+\nu \textbf{P}^\uparrow_{\rm GM}+(1-\mu-\nu)P^{\mu=\nu=0}_{SvL}, \label{3dfact}
\ee
where $\textbf{P}^\uparrow_{\rm GM}$ is a dominant Mermin box which satisfies a decomposition in terms of a $P^{mm}_M$ and $12$ different $P^{nmm}_M$'s
and $P^{\mu=\nu=0}_{SvL}$ is a box which does not have the irreducible dominant Svetlichny-box and Mermin-box weights.
 As already mentioned in Result $1$ the coefficient $\mu$ is called the Svetlichny strength of the correlation $P$, and we call the coefficient $\nu$ the Mermin strength of the given correlation.
\end{result}
\begin{proof}
Note that any Svetlichny-local box with SS equals to zero can be decomposed in a convex mixture of a dominant Mermin-box and
a box which does not have any irreducible dominant Svetlichny-box and Mermin-box components as in Eq. (\ref{mu0can}).
Thus by decomposing the Svetlichny-local box in the canonical decomposition (\ref{canoG>})
as in Eq. (\ref{mu0can}), we get the decomposition of the form given in Eq. (\ref{3dfact}) with
$\nu=\zeta(1-\mu)$.
\end{proof}

\begin{proposition}
If a box has $\mathcal{Q}=0$, then it necessarily has Mermin strength $\nu=0$. 
\end{proposition}
\begin{proof}
Notice that for any $P$ as given by the decomposition (\ref{3dfact}), MS $\nu=(1-\mu)\mathcal{Q}\left(\sum^{15}_{i=0} w_i P^i_{GM}\right)/4$, where $\sum^{15}_{i=0} w_i P^i_{GM}$
is the term that appears in the decomposition of $P^{\mu=0}_{SvL}$ as given in Eq. (\ref{stq1}). 
Therefore, boxes with $\mathcal{Q}=0$ necessarily have MS $\nu=0$.  
\end{proof}

It should be noted that there can be boxes which do not have MS, but nevertheless, have $\mathcal{Q}>0$.
For those boxes which lie on the line segment joining a Mermin-box and a Svetlichny-nonlocal/local box with 
$\mathcal{Q}=0$, the quantities $\mathcal{Q}$ and  MS are related to each other as follows.     
\begin{proposition}
For any given box $P\in \mathcal{R}$ which has a decomposition of the form,
\be
P=\mu \textbf{P}^\uparrow_{\rm Sv}+\nu \textbf{P}^\uparrow_{\rm GM}+(1-\mu-\nu)P^{\mu=\mathcal{Q}=0}_{SvL}, \label{3dfact1}
\ee
the quantities $\mathcal{Q}$ defined in Eq. (\ref{GMD}) and the MS satisfy the relationship 
$\mathcal{Q}(P)=4\nu$.
\end{proposition}
\begin{proof}
We rewrite any box $P\in\mathcal{R}$ as given by the decomposition (\ref{3dfact1}) as the convex mixture
\be
P=\nu \textbf{P}^\uparrow_{\rm GM}+(1-\nu)P_{\mathcal{Q}=0},\label{Qcan}
\ee
where $P_{\mathcal{Q}=0}:=\frac{1}{1-\nu}\left((1-\mu-\nu)P^{\mu=\mathcal{Q}=0}_{SvL}+\mu \textbf{P}^\uparrow_{\rm Sv}\right)$
which is a Svetlichny-nonlocal/local box with $\mathcal{Q}=0$.
Note that the set of $\mathcal{Q}=0$ boxes is nonconvex, i.e.,
certain convex combination of the $\mathcal{Q}=0$ boxes can have $\mathcal{Q}>0$.
Also there are $\mathcal{Q}=0$ boxes which
can be written as a convex mixture of two $\mathcal{Q}>0$ boxes. Thus, $\mathcal{Q}$ is not linear for these two types of decomposition.
However, $\mathcal{Q}$ is linear for the decomposition as given in Eq. (\ref{Qcan}) since the convex mixture of a $\mathcal{Q}>0$ box and a $\mathcal{Q}=0$ box is always a $\mathcal{Q}>0$ box.
Therefore, the box (\ref{Qcan})
has $\mathcal{Q}(P)=\nu \textbf{P}^\uparrow_{\rm GM} +(1-\nu)\mathcal{Q}(P_{\mathcal{Q}=0})=4\nu$.
\end{proof}

\begin{proposition}
The Mermin strength $\nu$ of any box is invariant under LRO and permutations of the parties. 
\end{proposition}
\begin{proof}
The statement of this proposition follows from the fact that $\mathcal{G}$ and $\mathcal{Q}$ are invariant under LRO
and permutations of the parties.  
\end{proof}

 With the quantities SS and MS, in the following section we will study the nonclassicality present in two-way local and fully local correlations in the restricted local Hilbert-space dimensional quantum simulation of those correlations. 
\section{Nonclassicality beyond the nonlocality paradigm}\label{QC}
Here we will show that for simulating a tripartite correlation having either nonzero SS or nonzero MS by a three qubit quantum state, it requires that the state must possess genuine nonclassicality either in the form of genuine discord \cite{GTC,GCTS,GTQD} or in the form of genuine entanglement. Before presenting the most general case, we first discuss two pure case scenarios.
  
\subsection{GHZ-class states}
The GHZ-class states which have bipartite entanglement between $A$ and $B$ are given as follows:
\be
|\psi_{gs}\rangle=\cos\theta|000\rangle+\sin\theta|11\rangle\Big\{\cos\theta_3|0\rangle+\sin\theta_3|1\rangle\Big\}\;.
\label{gs}
\ee
For this class of states the genuine tripartite entanglement is quantified by the three-tangle \cite{CKW}, $\tau_3=(\sin2\theta\sin\theta_3)^2$, and the bipartite entanglement is quantified by the concurrence \cite{WKW}, $C_{12}=\sin2\theta\cos\theta_3$. To study the nonclassicality of boxes arising from this class of states, we restrict ourselves to projective measurements, i.e., Alice, Bob, and Charlie perform projective measurements denoted by $\hat{a}_i.\vec{\sigma}$, $\hat{b}_j.\vec{\sigma}$, and $\hat{c}_k.\vec{\sigma}$ respectively on their subsystems. Here, $\hat{a}_i$, $\hat{b}_j$, and $\hat{c}_k$ are unit Bloch vectors denoting the measurement directions and $\vec{\sigma}=\{\sigma_1,\sigma_2,\sigma_3\}$, with $\{\sigma_i\}_{i=1,2,3}$ being the Pauli matrices and $i,j,k\in\{0,1\}$. 

\emph{Svetlichny nonclassicality}: Let us choose the following particular measurements:
$
\hat{a}_0\!=\!\hat{x},~ \hat{a}_1\!=\!\hat{y};~
\hat{b}_j\!=\!\frac{1}{\sqrt{2}}\left(\hat{x}+(-1)^{j\oplus 1}\hat{y}\right);~ \hat{c}_0\!=\!\hat{x},~ \hat{c}_1\!=\!\hat{y}.$
It can be checked that for this particular set of measurements, boxes arising from the GHZ-class states 
satisfy a decomposition of the following form:
\be
P=\frac{\sqrt{\tau_3}}{\sqrt{2}}P^{0000}_{Sv}+\left(1-\frac{\sqrt{\tau_3}}{\sqrt{2}}\right)P^{\mathcal{G}=0}_{SvL}. \label{optSDdec}
\ee
The box $P^{\mathcal{G}=0}_{SvL}$ which has nonmaximally mixed marginals becomes white noise for $\theta_3=\pi/2$ (the GGHZ states). Also tedious but straightforward calculation gives $\mathcal{G}=4\sqrt{2\tau_3}$ for these boxes.  The correlations (\ref{optSDdec}) 
exhibit Svetlichny nonlocality when the three-tangle $\tau_3>1/2$ and for $\tau_3\le1/2$ it has no genuine nonlocality. Also note that these correlations are two-way local for $C_{12}>1/\sqrt{2}$ and $\tau_3\le1/2$. However for $0<\tau\le 1$, these correlations are Svetlichny nonclassical since they have nonzero SS.

\emph{Mermin nonclassicality}: Let us choose the following particular measurements:
$
\hat{a}_0=\hat{b}_0=\hat{c}_0=\hat{x}, \quad \hat{a}_1=\hat{b}_1=\hat{c}_1=\hat{y}.
$
Boxes arising from this set of measurements on the GHZ-class states belong to $\mathcal{L}_2$ and they have $\mathcal{G}=0$ and $\mathcal{Q}=4\sqrt{\tau}_3$ and it has a decomposition of the form given in Result $2$ as follows: 
\ba
P&=&\sqrt{\tau_3}P_{GM}+\left(1-\sqrt{\tau_3}\right)P^{\mathcal{G}=\mathcal{Q}=0}_{L}\nonumber\\
&=&\sqrt{\tau_3}\left(\frac{P^{0000}_{Sv}+P^{1110}_{Sv}}{2}\right)+\left(1-\sqrt{\tau_3}\right)P^{\mathcal{G}=\mathcal{Q}=0}_{L}. \label{MDGHZop}
\ea
The box $P^{\mathcal{G}=\mathcal{Q}=0}_{L}$ which in this case is a fully local box and has nonmaximally mixed marginals in general becomes white noise for the GGHZ states. These correlations are nonlocal when the three-tangle $\tau_3>1/4$, otherwise they are fully local. However for $0<\tau\le 1$, these correlations are Mermin nonclassical since they have nonzero MS.

In Ref. \cite{JebaEPR}, the author has derived two tripartite steering inequalities to detect
two types of genuine steering which are called Svetlichny steering and Mermin steering. It was demonstrated that
the correlations which exhibit Svetlichny or Mermin steering detect genuine entanglement of $\mathbb{C}^2\otimes\mathbb{C}^2\otimes\mathbb{C}^d$ systems.
It is noted that the boxes (\ref{optSDdec})  exhibit Svetlichny steering
for SS $\mu>1/4$, whereas the boxes (\ref{MDGHZop}) exhibit Mermin steering for MS $\nu>1/4$.   
This implies that in this range, genuine entanglement is necessary for simulating these correlations by using
$\mathbb{C}^2\otimes\mathbb{C}^2\otimes\mathbb{C}^2$ systems, even if the boxes (\ref{optSDdec}) do not violate the SI and 
the boxes (\ref{MDGHZop}) do not violate the MI more than $2\sqrt{2}$.

It is noted that there are correlations arising from the GHZ-class states which exhibit genuine nonlocality,
but nevertheless, have SS $\mu=0$ (see Appendix \ref{limSSMS}). 

\emph{Svetlichny-Mermin nonclassicality}: For the state dependent measurement setting 
$\hat{a}_0=\hat{x}$, $\hat{a}_1=\hat{y}$;
$\hat{b}_0=\hat{x}\sin2\theta-\hat{y}\cos2\theta$,
$\hat{b}_1=\hat{x}\cos2\theta+\hat{y}\sin2\theta$;
$\hat{c}_0=\hat{x}$, $\hat{c}_1=\hat{y}$, we find that the correlations arising from the GGHZ states (i.e., states with $\theta_3=\pi/2$ in Eq. (\ref{gs})) admit a decomposition of the form given in Result $2$:
\be
P=\frac{\mathcal{G}}{8} P^{0000}_{Sv}+\frac{\mathcal{Q}}{4} \left(\frac{P^{0000}_{Sv}+P^{111\gamma}_{Sv}}{2}\right)+\left(1-\frac{\mathcal{G}}{8} -\frac{\mathcal{Q}}{4} \right)P_N.\label{cdQC}
\ee
Here,
\ba
\mathcal{G}&=&\left\{\begin{array}{lr}
8\tau_3 \quad \text{when} \quad 0 \le \theta \le \frac{\pi}{8}\\
8\sqrt{\tau_3(1-\tau_3)}  \quad \text{when} \quad \frac{\pi}{8} \le \theta \le \frac{\pi}{4},\\
\end{array}
\right.\nonumber\\
\mathcal{Q}&=&4\left|\tau_3-\sqrt{\tau_3(1-\tau_3)}\right| \nonumber
\ea
and $P_N$ is the white noise which has $\mathcal{G}=\mathcal{Q}=0$. In Eq. (\ref{cdQC}), $\gamma=0$ for $0 \le \theta \le \frac{\pi}{8}$ and $\gamma=1$ otherwise.

\subsection{W-class states}
Consider the W-class states given by the following form:
\be
\ket{\psi_w}=\mu_1\ket{100}+\mu_2\ket{010}+\mu_3\ket{001}, \label{W}
\ee
where $\mu_i\ge 0$ and $\sum_{i=1}^3\mu_i^2=1$. These states have the three-tangle $\tau_3=0$ and have all the three bipartite concurrences nonzero:
$C_{12}=2\mu_1\mu_2$, $C_{13}=2\mu_1\mu_3$ and $C_{23}=2\mu_2\mu_3$. The W-class states have
\emph{minimal concurrence of assistance} \cite{Chietal}
$C^a_{min}:=\min\{C_{12},C_{13},C_{23}\}$, which can be considered as a genuine tripartite entanglement measure of this class of states. Interestingly, we find a relationship between nonzero Svetlichny/Mermin strength of the boxes arising from the W-class states and $C^a_{min}$.

\emph{Svetlichny nonclassicality}: Consider the following measurements:
$
\hat{a}_0\!=\!\hat{z},~ \hat{a}_1\!=\!\hat{x};~
\hat{b}_j\!=\!\frac{1}{\sqrt{2}}\left(\hat{z}+(-1)^{j\oplus 1}\hat{x}\right);~ \hat{c}_0\!=\!\hat{z},~ \hat{c}_1\!=\!\hat{x}. $
For this setting the correlations arising from the W-class states violate the Svetlichny
inequality when $C_{12}+C_{13}+C_{23}>2\sqrt{2}-1$, showing Svetlichny nonlocality and hence Svetlichny nonclassicality \footnote{At this point we wish to draw the attention to an interesting observation. Correlation obtained from some particular state and particular setting showing optimal Svetlichny nonlocality may not show optimal Svetlichny nonclassicality for the same state with same setting. We calculate $\mathcal{G}$ for the optimal Svetlichny nonlocal setting for the W-class provided in Ref. \cite{Ajoy} and find that it is less than that of the correlations in Eq. (\ref{WclassSD}).}.  We find that these correlations satisfy a decomposition of the following form:
\be
P=\frac{C^a_{min}}{\sqrt{2}}P^{0100}_{Sv}+\left(1-\frac{C^a_{min}}{\sqrt{2}}\right)P^{\mathcal{G}=0}_{SvL}. \label{WclassSD}
\ee
Straightforward calculation for this correlation gives $\mathcal{G}=4\sqrt{2}C^a_{min}$ which can be nonzero (hence have Svetlichny nonclassicality) even when it does not violate a SI.

\emph{Mermin nonclassicality}: Let us choose the following set of measurements:
$
\hat{a}_0=\hat{b}_0=\hat{c}_0=\hat{z}, \quad \hat{a}_1=\hat{b}_1=\hat{c}_1=\hat{x}.\label{MDxy}$
Correlations arising from this setting for the W-class states exhibit Mermin nonlocality \emph{iff} $C_{12}+C_{13}+C_{23}>1$. For these correlations it turns out that $\mathcal{Q}=4C^a_{min}$ and  the correlations can be decomposed as
\be
P=C^a_{min}\left[\frac{P^{0001}_{Sv}+P_{Sv}^{1111}}{2}\right]+\left(1-C^a_{min}\right)P^{\mathcal{Q}=0}_{SvL}. \label{WclassMD}
\ee
In this case we have $P^{\mathcal{Q}=0}_{SvL}\in\mathcal{L}$. As shown for the GHZ-class, here also one can find example of boxes arising from W-class states having $\mathcal{G}\ne 0\ne\mathcal{Q}$.

All the examples discussed above establish the fact that if the correlations arising from three-qubit systems have nonzero SS or 
nonzero MS, then the states would necessarily have genuine quantumness. In what follows, we present our main result that shows the converse, i.e., three-qubit states with no genuine quantumness can never give rise to correlations having either nonzero SS or nonzero MS.    
\subsection{Three-qubit states without genuine quantumness have $SS=MS=0$}
Before presenting our result, we first briefly discuss about the genuine quantumness of tripartite states. Quantumness in states is a vastly studied issue in quantum information theory and commonly captured by the quantity entanglement \cite{Horodecki}. However, there exist unentangled quantum states that give rise to advantage for certain bipartite quantum information tasks \cite{QSM,QDRS}. This uphold the study of quantumness in states that goes beyond the standard regime of entanglement. In the bipartite scenario \emph{quantum discord}, introduced in \cite{WZQD} (see also \cite{Modietal}), is one way to capture such quantumness in states. In the multipartite scenario,  recently Giorgi \etal \cite{GTC} generalized the notion of quantumness and they defined genuine multipartite quantum discord. The genuine multipartite quantum discord estimates the quantum part of genuine multipartite correlations.

A bipartite quantum state shared by Alice and Bob has zero quantum discord  ($A$ to $B$) \emph{iff} it cannot be written in the form of classical-quantum (CQ) state, i.e., $\rho_{CQ}=\sum_ip_i|i\rangle^{A}\langle i| \otimes \rho^B_i$, where $\{|i\rangle\}_i$ is some orthonormal basis of Alice's Hilbert space $\mathcal{H}_A$ and $\rho_i$ are arbitrary quantum states \cite{CQstate,Dakicetal}.  Quantum-classical (QC) states have similar form but with permutation of the parties' indices $A$ and $B$. Bipartite states which are neither CQ nor QC are called quantum-quantum (QQ) states.

A tripartite state $\rho_{bs}$ is biseparable  if it can be written as a convex mixture of the states that are separable with respect
to some partition, i.e.,
\be
\rho_{bs}=\sum_k p_k \rho_k^A \otimes \rho_k^{BC}+\sum_k q_k \rho_k^{AB} \otimes \rho_k^{C}+\sum_k r_k \rho_k^{AC} \otimes \rho_k^{B},
\ee
with  $\sum_k p_k+\sum_k q_k+\sum_k r_k=1$. 
Let us consider the set $S_0:=\left\{\rho_{sep}^{A|BC}, \rho_{sep}^{AB|C}, \rho_{sep}^{AC|B}\right\}$ of the three-qubit states, where
\be
\rho_{sep}^{A|BC}=\sum_ip_i \rho^A_i \otimes \rho^{BC}_i, \label{BC}
\ee
with the bipartite marginals $\mathrm{Tr}_{B}\left[\rho^{A|BC}_{sep}\right]$ and $\mathrm{Tr}_{C}\left[\rho^{A|BC}_{sep}\right]$ are either a CQ state or a QC state;  and $\rho_{sep}^{AB|C}$ and $\rho_{sep}^{AC|B}$ are defined analogously. States which are not in this form must have quantum correlation shared among all three subsystems. In Ref. \cite{GTC}, Giorgi \etal defined \emph{genuine tripartite quantum discord} to capture the quantumness shared among all the parties. The fully separable 
states which are CQ states with respect to the bipartite cut A versus BC can be decomposed as
\be
\rho^{A|BC}_{CQ}=\sum_ip_i |i\rangle^{A}\langle i| \otimes \rho^{B}_i \otimes \rho^{C}_i, \label{cqBC}
\ee
where $\{|i\rangle^{A}\}_i$ is some orthonormal basis of Alice's Hilbert space $\mathcal{H}_A$. The QC states
$\rho^{AB|C}_{QC}$ and  $\rho^{AC|B}_{QC}$ with respect to the other two bipartite cuts are defined similarly.
The CQ and QC states defined above do not have nonzero genuine quantum discord since one of the subsystems is always classically correlated with the other two subsystems. Notice that the set of CQ and QC states forms a proper subset of the set $S_0$
defined above. Let us denote $S_{NG}$ as the set of all biseparable states which do not have genuine tripartite quantum correlations and we have that $S_{NG}\subseteq S_0$. 
We will now demonstrate that the three-qubit states that belong to the set $S_0$ cannot be used to demonstrate nonzero SS or nonzero MS.

A multipartite quantum state is said to possess genuine multipartite correlations when it is nonproduct in every bipartite cut \cite{FCBox}.
From this perspective, genuine multipartite quantum discord was defined to quantify genuine multipartite quantum correlations in quantum states \cite{GTC,GCTS,GTQD}.
Analogously, we say that a tripartite box possesses genuine tripartite correlations when it cannot be written as a product of marginals corresponding to any two parties and the third party, i.e., 
$P(abc|xyz)\ne P(a|x)P(bc|yz)$, and permutations.
It is noted that the boxes which have nonzero SS or nonzero MS have genuine tripartite correlations.
Therefore, any biseparable state of the form $\rho=\rho_{AB}\otimes \rho_C$ or any other permutation 
can never give rise to nonzero SS or nonzero MS for all possible measurements.

We are  now interested in mixed three-qubit states $\rho\in S_0$ which have genuine tripartite correlations.  
Consider the states,
\be
\rho_{sep}^{AB|C}=\sum_iq_i \rho^{AB}_i \otimes \rho^{C}_i, \label{AB}
\ee
which are the permutation of the states as given in Eq. (\ref{BC}).
It is noted that in the case of pure genuinely entangled states,
optimal nonzero SS or nonzero MS is achieved when the number of nonzero Svetlichny functions $\mathcal{S}_{\alpha\beta\gamma}$ or Mermin functions $\mathcal{M}_{\alpha\beta\gamma}$ is minimized  over all possible projective measurements that give rise to $\mathcal{G}>0$ or $\mathcal{Q}>0$.
Thus, for the purpose of demonstrating that the states (\ref{AB}) cannot give rise to nonzero SS or nonzero MS,  
we will now check whether these states give rise to $\mathcal{G}>0$ or $\mathcal{Q}>0$ for the projective measurements that maximize minimal number of nonzero Svetlichny functions or nonzero Mermin functions. 
For the states (\ref{AB}), the expectation value factorizes as follows:
\be
\braket{A_xB_yC_z}=\sum_i q_i \braket{A_xB_y}_i\braket{C_z}_i. \label{exfact}
\ee
This implies that the quantity $\mathcal{G}_1$ appearing in the right-hand side of Eq. (\ref{GBD}) factorizes as follows:
\begin{widetext}
\begin{align}
\mathcal{G}_1&=|\Big|\!\left|\sum_iq_i \left\{\mathcal{B}^i_{000}\braket{C_0}_i+\mathcal{B}^i_{111}\braket{C_1}_i\right\}\right|-
\left|\sum_iq_i\!\left\{\mathcal{B}^i_{000}\braket{C_0}_i-\mathcal{B}^i_{111}\braket{C_1}_i\right\}\right|\Big|\nonumber \\
&-\Big|\left|\sum_iq_i\left\{\mathcal{B}^i_{000}\braket{C_1}_i\!+\mathcal{B}^i_{111}\braket{C_0}_i\right\}\right|
-\left|\sum_iq_i\left\{\mathcal{B}^i_{000}\braket{C_1}_i-\mathcal{B}^i_{111}\braket{C_0}_i\right\}\right|\Big|\nonumber\\
&-|\Big|\left|\sum_iq_i \left\{\mathcal{B}^i_{010}\braket{C_0}_i+\mathcal{B}^i_{100}\braket{C_1}_i\right\}\right|-
\left|\sum_iq_i\!\left\{\mathcal{B}^i_{010}\braket{C_0}_i-\mathcal{B}^i_{100}\braket{C_1}_i\right\}\right|\Big|\nonumber\\
&-\Big|\left|\sum_iq_i\!\left\{\mathcal{B}^i_{010}\braket{C_0}_i+\mathcal{B}^i_{100}\braket{C_1}_i\right\}\right|-
\left|\sum_iq_i\left\{\mathcal{B}^i_{010}\braket{C_1}_i-\mathcal{B}^i_{100}\braket{C_0}_i\right\}\right|\Big||. \label{ExQC}
\end{align}
\end{widetext}
Here, $\mathcal{B}^i_{\alpha\beta\gamma}$ are the Bell-CHSH operators having the following expressions \cite{chsh,WernerWolf}:
\ba
\mathcal{B}_{\alpha\beta\gamma}&:= &(-1)^\gamma\braket{A_0B_0}+(-1)^{\beta \oplus \gamma}\braket{A_0B_1}\nonumber\\
&+&(-1)^{\alpha \oplus \gamma}\braket{A_1B_0}+(-1)^{\alpha \oplus \beta \oplus \gamma \oplus 1} \braket{A_1B_1}. \nonumber \label{BCHSH}
\ea
In Eq. (\ref{ExQC}), $\mathcal{B}^i_{\alpha\beta\gamma}$ and $\braket{C_z}_i$ are evaluated for $\rho^i_{AB}$ and $\rho^i_C$
given in Eq. (\ref{AB}).   
Let us now try to maximize  minimal number of nonzero Svetlichny functions with respect to $\rho_{sep}^{AB|C}$ in which $\rho^i_{AB}$ are  QQ states.
Since the four Bell-CHSH operators $\mathcal{B}^i_{\alpha\beta\gamma}$ in Eq. (\ref{ExQC}) are evaluated for the QQ states, we can find suitable projective measurements on Alice and Bob's sides that give rise to nonzero for only one of  $\mathcal{B}^i_{\alpha\beta\gamma}$ \cite{Jeba1,Jeba}.
For some measurements on Alice and Bob's sides that have this property, it is readily seen that
$\mathcal{G}_1=0$ for any choice of measurements on Charlie's side. Therefore, we can conclude that
the biseparable states $\rho_{sep}^{AB|C}$ do not give rise to nonzero SS for any choice of projective measurements.

To give evidence for the above claim, we consider the following biseparable states which have genuine tripartite correlations: 
\begin{align}
\rho&=p|\psi^+\rangle^{AB}\langle \psi^+|\otimes|0\rangle^{C}\langle 0| +(1-p)|\psi^-\rangle^{AB}\langle \psi^-|\otimes|+_x\rangle^{C}\langle +_x|,
\end{align}
where $p\ge1/2$, $\ket{\psi^+}=\frac{1}{\sqrt{2}}(\ket{00}+\ket{11})$, $\ket{\psi^-}=\frac{1}{\sqrt{2}}(\ket{00}-\ket{11})$
and $\ket{+_x}=\frac{1}{\sqrt{2}}(\ket{0}+\ket{1})$. For the above state and for the measurements $\hat{a}_0\!=\!\hat{x},~ \hat{a}_1\!=\!\hat{y};~
\hat{b}_j\!=\!\frac{1}{\sqrt{2}}\left(\hat{x}+(-1)^{j\oplus 1}\hat{y}\right)$ on Alice and Bob's sides, it can be checked that
the four Svetlichny functions are maximized and the other four are zero and these values give $\mathcal{G}_1=0$.

Similarly, it can be demonstrated that MS cannot be nonzero for the states $\rho_{sep}^{AB|C}$ as given in Eq. (\ref{AB}) in which $\rho^i_{AB}$ are QQ states
by exploiting the factorization property in Eq. (\ref{exfact}).
Since $\mathcal{G}$ and $\mathcal{Q}$ are symmetric under the permutations of the parties, the states $\rho_{sep}^{A|BC}$
and $\rho_{sep}^{AC|B}$ which have genuine tripartite correlations and quantum correlations between two subsystems do not give rise to $\mathcal{G}>0$ or $\mathcal{Q}>0$ for the above type of measurement scenario. 
Thus, we can state the following our main result: 
\begin{result}
When the measured quantum state is restricted to be that of a three-qubit system,  genuine quantumness as captured by the notion of genuine quantum discord is necessary for demonstrating nonzero SS or nonzero MS for projective measurements.
\end{result}

\begin{remark}
Suppose a given biseparable or fully separable state can give rise to nonzero SS or nonzero MS.
Then the state cannot be written in the form of the separable states in the set $S_0$ and has a decomposition
as follows:
\be
\rho\!=\!t_1\rho_{sep}^{A|BC} +t_2 \rho_{sep}^{AB|C} +t_3\rho_{sep}^{AC|B}, \label{GTQC}
\ee
which is a mixture of the separable states in $S_0$ with more than one coefficient $t_1$, $t_2$, $t_3$ nonzero.
\end{remark}

As an illustration, we consider a fully separable state which is neither a CQ state nor a QC state. The state is given as follows:
\begin{eqnarray}
\rho&=&\frac{1}{4}\{\Pi_x^{A} \otimes \Pi_x^{B} \otimes \Pi_x^{C} +\Pi_x^{A} \otimes \Pi_y^{B} \otimes \Pi_y^{C}\nonumber \\
&+&\Pi_y^{A} \otimes \Pi_y^{B} \otimes \Pi_x^{C} +\Pi_y^{A} \otimes \Pi_x^{B} \otimes \Pi_y^{C}\}, \label{QQstate}
\end{eqnarray}
where $\Pi_x=\ketbra{+_x}{+_x}$ and $\Pi_y=\ketbra{+_y}{+_y}$ are the eigenstates of $\sigma_x$ and $\sigma_y$, respectively.
For the measurements that give rise to the box (\ref{optSDdec}), the above state gives rise to the isotropic Svetlichny-box  (\ref{SvF}) with SS $\mu=1/4$. 
Note that the state (\ref{QQstate}) admits a decomposition of the form given in Eq. (\ref{GTQC})
as follows:
\begin{eqnarray}
\rho&=&\frac{1}{6}\{\Pi_x^{A} \otimes \rho_1^{BC}+\Pi_y^{A} \otimes  \rho_2^{BC}+ \rho_1^{AB} \otimes \Pi_x^{C}
+\rho_2^{AB} \otimes \Pi_y^{C} \nonumber \\
&+& \rho_1^{AC}\otimes \Pi_x^{B}+ \rho_2^{AC}\otimes \Pi_y^{B} \},
\end{eqnarray}
where $\rho_1^{BC}:=(\Pi_x^{B} \otimes \Pi_x^{C} + \Pi_y^{B} \otimes \Pi_y^{C})/2$, $\rho_2^{BC}:=(\Pi_x^{B} \otimes \Pi_y^{C} + \Pi_y^{B} \otimes \Pi_x^{C})/2$, $\rho_1^{AC}:=(\Pi_x^{A} \otimes \Pi_x^{C} + \Pi_y^{A} \otimes \Pi_y^{C})/2$, $\rho_2^{AC}:=(\Pi_x^{A} \otimes \Pi_y^{C} + \Pi_y^{A} \otimes \Pi_x^{C})/2$, $\rho_1^{AB}:=(\Pi_x^{A} \otimes \Pi_x^{B} + \Pi_y^{A} \otimes \Pi_y^{B})/2$ and  $\rho_2^{AB}:=(\Pi_x^{A} \otimes \Pi_y^{B} + \Pi_y^{A} \otimes \Pi_x^{B})/2$ which are all
QQ states.

\section{Conclusions}\label{conc}
In this work, we have studied quantum simulation of tripartite correlations with restricted local Hilbert-space dimensions, in particular, we have considered $\mathbb{C}^2\otimes\mathbb{C}^2\otimes\mathbb{C}^2$ quantum systems. 
We have proposed two quantities of interest, called \emph{Svetlichny strength} and \emph{Mermin strength},
which capture genuine nonclassicality in correlations beyond the standard nonlocality paradigm.
We find that in the restricted simulation scenario, fully local and two-way local correlations having nonzero Svetlichny strength or nonzero Mermin strength must require the presence of \emph{genuine quantumness} in the three-qubit states that are used to simulate those correlations. For the Svetlichny-box polytope which is a proper subpolytope of full tripartite nonsignaling polytope, we have derived a canonical decomposition and characterized the scenarios how genuine nonclassicality is manifested in the correlations of these canonical form.

This work motivates an interesting question for future research. In Ref. \cite{JebaEPR}, the author has demonstrated that  
the tripartite correlations which exhibit Svetlichny steering or Mermin steering detect genuine entanglement in a semi-device-independent way (i.e., by assuming only the local Hilbert-space 
dimensions and without any assumptions on the measurements performed). Similarly, it would be interesting to show that the two-way local and fully local correlations that have nonzero Svetlichny strength or nonzero Mermin strength detect genuine quantum discord in a semi-device-independent way.  

\section*{Acknowledgements}
I wish to thank IMSc., Chennai  and
IISER Mohali for financial support, and
Dr. Pranaw Rungta for suggesting this problem and for
many inspiring discussions subsequently.
I am thankful to Dr. R. Srikanth and Dr. Manik Banik for valuable suggestions and comments, 
and Dr. K. P. Yogendran, Dr. Sibasish Ghose, Dr. Prabha Mandayam and Dr. Arul Lakshminarayan for discussions.
I acknowledge support through the project SR/S2/LOP-08/2013 of the DST,
Govt.  of India.

\appendix

\section{Derivation of Result $2$}
\begin{Lemma}
A two-way local box that has $\mathcal{Q}=4$ is, in general,
a convex combination of a $P^{mm}_M$ and $12$ different $P^{nmm}_{M}$'s, i.e., 
\be
P_{GM}=uP^{mm}_M+\sum^{12}_{i=1}v_iP^{nmm}_{M_i}.
\ee
All the Mermin-boxes in this decomposition violate the same Mermin inequality as they all have same values for $\braket{A_xB_yC_z}$.
\end{Lemma}
\begin{proof}
Notice that any convex mixture of the two Mermin boxes given in Eqs. (\ref{Mbdec}) and (\ref{Mbnmm}) has $\mathcal{Q}=4$.
It can be checked that there can be four Mermin boxes which are the uniform mixture of two bipartite PR-boxes in
Eq. (\ref{Mbdec}). The permutation of the parties' indices in these boxes give rise to the another $8$ Mermin boxes.
Since these $12$ Mermin boxes are equivalent with respect to $\braket{A_xB_yC_z}$,  any convex mixture of these $12$ Mermin boxes and the Mermin box (\ref{Mbdec}) again has
$\mathcal{Q}=4$. Similarly, any other maximally two-way local box with $\mathcal{Q}=4$ is, in general, a convex mixture of the
corresponding $13$ Mermin boxes. 
\end{proof}

\begin{Lemma}
Any Svetlichny-local box, $P^{\mu=0}_{SvL}\in\mathcal{L}_2$, which does not have an irreducible dominant Svetlichny-box weight
can be written as the mixture
\be
P^{\mu=0}_{SvL}=\zeta \textbf{P}^\uparrow_{\rm GM} +(1-\zeta)P^{\mu=\nu=0}_{SvL}, \label{mu0can}
\ee
where $P^{\mu=\nu=0}_{SvL}$ is a Svetlichny-local box with irreducible
dominant Svetlichny-box and Mermin-box weights equal to zero. 
\end{Lemma}  
\begin{proof}
Since the Mermin-boxes belong to the two-way local polytope,
we can express any box $P^{\mu=0}_{SvL}\in\mathcal{L}_2$ as a convex mixture of the Mermin-boxes and a Svetlichny-local box 
that does not have any Mermin-box component,
\be
P^{\mathcal{\mu}=0}_{SvL}=\sum^{15}_{i=0} u_iP^{mm}_{M_i}+\sum^{192}_{j=1}v_jP_{M_j}^{nmm}+\left(1-\sum^{15}_{i=0}u_i-\sum^{192}_{j=1}v_j\right)P_{SvL}. \label{g=0}
\ee
It follows from Lemma $1$ that the mixture of the Mermin boxes in Eq. (\ref{g=0}) can be reexpressed as
the mixture of the $16$ different $P_{GM}$'s that have $\mathcal{Q}=4$. Therefore, we
simplify the decomposition (\ref{g=0}) as follows:
\be
P^{\mathcal{\mu}=0}_{SvL}=\sum^{15}_{i=0}w_iP^i_{GM}+\left(1-\sum^{15}_{i=0}u_i-\sum^{192}_{j=1}v_j\right)P_{SvL}. \label{stq1}
\ee

We now express the first term in the above decomposition as a sum of the local boxes, $P^i_{L}$,
and a \emph{residual} box, $\textbf{P}^\uparrow_{\rm GM}$ (the dominant Mermin-box), i.e.,
\be
\sum^{15}_{i=0}w_iP^i_{GM}=\zeta \textbf{P}^\uparrow_{\rm GM}+\sum^{15}_{i=1} l_iP^i_{L}. \label{stq2}
\ee
This is done by recursively rewriting the mixture of two different Mermin-boxes 
as the mixture of the single Mermin-box and a local box, i.e., $pP^i_{GM}+qP^j_{GM}= (p-q)P^i_{GM}+2qP_L$. Here, $p>q$ and $P_L:=\frac{1}{2}\left(P^i_{GM}+P^j_{GM}\right)$ which is a local box. 
Since finding the above residual decomposition is not unique, $\zeta$ in Eq. (\ref{stq2}) can take different value for different residual decomposition.

The evaluation of $\mathcal{Q}_1$ for the mixture of the Mermin-boxes gives,  
\begin{align}
&\mathcal{Q}_1\left(\sum^{15}_{i=0} w_i P^i_{GM}\right)\nonumber \\
&=4\Big||\Big||w_0-w_1|-|w_2-w_3|\Big|-\Big||w_4-w_5|-|w_6-w_7|\Big||\nonumber \\
&-|\Big||w_8-w_9|-|w_{10}-w_{11}|\Big|-\Big||w_{12}-w_{13}|-|w_{14}-w_{15}|\Big||\Big|.
\end{align} 
Thus, $\mathcal{Q}\left(\sum^{15}_{i=0} w_i P^i_{GM}\right)/4$ gives the (minimal) weight of the irreducible Mermin-box that dominates
over a local box.
For this reason, among all residual decompositions as in Eq. (\ref{stq2})
we take that particular decomposition where $\zeta=\mathcal{Q}\left(\sum^{15}_{i=0} w_i P^i_{GM}\right)/4$. 
Plugging this minimal residual decomposition in Eq. (\ref{stq1}), we get the following decomposition:
\be
P^{\mu=0}_{SvL}=\zeta \textbf{P}^\uparrow_{\rm GM}+(1-\zeta)P'_{SvL}, \label{g=0cano}
\ee
where $P'_{SvL}=\frac{1}{1-\zeta}\left\{\sum^{15}_{i=1}l_iP^{i}_L+\left(1-\sum^{15}_{i=0}u_i-\sum^{192}_{j=1}v_{j}\right)P_{SvL}\right\}$.
It is obvious that this box does not have any nonzero irreducible dominant Mermin-box weight since
the maximal irreducible Mermin-box weight is isolated in $\zeta$. 
\end{proof}

\section{Limitations of SS and MS as measures of genuine nonclassicality} \label{limSSMS}
Bancal \etal \cite{Banceletal} conjectured that all genuinely entangled pure states can give rise to three-way nonlocal correlations.
Bancal \etal noticed that there are three-way nonlocal correlations arising from the pure three-qubit states which do not violate a Svetlichny inequality.
This implies that these quantum boxes do not belong to the three-way nonlocal region of the Svetlichny-box polytope.
In Ref. \cite{gnl99}, it was shown that all
the GGHZ states $\ket{\psi_{GGHZ}}=\cos\theta\ket{000}+\sin\theta\ket{111}$ can be used to demonstrate three-way nonlocality, which is witnessed by violation of the class $99$ facet inequality of $\mathcal{L}_2$ given in Ref. \cite{Banceletal}.
Consider  the following class $99$ facet inequality:
\be
\mathcal{L}^{99}_2=\braket{A_0B_0}+\braket{A_0C_0}+\braket{B_1C_0}+\braket{A_1B_0C_1}-\braket{A_1B_1C_1}\le3. \label{NSFI}
\ee
The violation of this inequality by the GGHZ states is achievable with the
measurement setting $\hat{a}_0=\hat{z}$, $\hat{a}_1=\hat{x}$;
$\hat{b}_j=\cos t\hat{z}+(-1)^{j}\sin t\hat{x}$;
$\hat{c}_0=\hat{z}$, $\hat{c}_1=\hat{x}$, where $\cos t=1/\sqrt{1+\sin^22\theta}$.
For this setting, the correlations arising from
the GGHZ states  have $\mathcal{L}_2^{99}=1+2\sqrt{1+\sin^22\theta}>3$ for any $\tau_3>0$.
For $\theta=\frac{\pi}{4}$ (the GHZ state), the correlation violates the class $99$ facet inequality to its quantum bound of $1+2\sqrt{2}$.
The box which gives this violation
can be written as a mixture of the class $8$ extremal three-way nonlocal box given in the table of Ref. \cite{Pironioetal} and a local box as follows:
\be
P=\frac{1}{\sqrt{2}} P_8+\left(1-\frac{1}{\sqrt{2}}\right)P_L. \label{GHZopGNL}
\ee
Here, $P_8$ is the extremal three-way nonlocal box that belongs to the class $8$ and
$P_L$ is the local box arising from the state $\rho=\rho_{AC}\otimes \frac{\one}{2}$, where $\rho_{AC}=\frac{1}{2}\left(\ketbra{00}{00}+\ketbra{11}{11}\right)$ and $\frac{\one}{2}$ is the maximally mixed state, for the
above measurements.
Since the three-way nonlocality of the above correlation is due to the class $8$ extremal box, it does not violate a Svetlichny inequality.
Therefore, the box in Eq. (\ref{GHZopGNL}) lies outside the Svetlichny-box polytope and has $\mathcal{G}=\mathcal{Q}=0$.

We also find that the correlations (\ref{WclassSD}) violate the  class $99$ facet inequality when ${C_{13}+\left(C_{12}+C_{23}\right)/\sqrt{2}}>3-\sqrt{2}$. Therefore the Svetlichny-local box part in Eq. (\ref{WclassSD}), i.e., $P^{\mathcal{G}=0}_{SvL}$ must have nonzero fraction of class $8$ extremal box. This implies that whenever ${C_{13}+\left(C_{12}+C_{23}\right)/\sqrt{2}}>3-\sqrt{2}$ the correlations (\ref{WclassSD}) lie outside Svetlichny-box polytope and therefore these are the example of correlations $P \notin\mathcal{R}$ but having nonzero SS. 

\bibliographystyle{apsrev4-1}
\bibliography{SDQDW}

\begin{thebibliography}{42}%
\makeatletter
\providecommand \@ifxundefined [1]{%
 \@ifx{#1\undefined}
}%
\providecommand \@ifnum [1]{%
 \ifnum #1\expandafter \@firstoftwo
 \else \expandafter \@secondoftwo
 \fi
}%
\providecommand \@ifx [1]{%
 \ifx #1\expandafter \@firstoftwo
 \else \expandafter \@secondoftwo
 \fi
}%
\providecommand \natexlab [1]{#1}%
\providecommand \enquote  [1]{``#1''}%
\providecommand \bibnamefont  [1]{#1}%
\providecommand \bibfnamefont [1]{#1}%
\providecommand \citenamefont [1]{#1}%
\providecommand \href@noop [0]{\@secondoftwo}%
\providecommand \href [0]{\begingroup \@sanitize@url \@href}%
\providecommand \@href[1]{\@@startlink{#1}\@@href}%
\providecommand \@@href[1]{\endgroup#1\@@endlink}%
\providecommand \@sanitize@url [0]{\catcode `\\12\catcode `\$12\catcode
  `\&12\catcode `\#12\catcode `\^12\catcode `\_12\catcode `\%12\relax}%
\providecommand \@@startlink[1]{}%
\providecommand \@@endlink[0]{}%
\providecommand \url  [0]{\begingroup\@sanitize@url \@url }%
\providecommand \@url [1]{\endgroup\@href {#1}{\urlprefix }}%
\providecommand \urlprefix  [0]{URL }%
\providecommand \Eprint [0]{\href }%
\providecommand \doibase [0]{http://dx.doi.org/}%
\providecommand \selectlanguage [0]{\@gobble}%
\providecommand \bibinfo  [0]{\@secondoftwo}%
\providecommand \bibfield  [0]{\@secondoftwo}%
\providecommand \translation [1]{[#1]}%
\providecommand \BibitemOpen [0]{}%
\providecommand \bibitemStop [0]{}%
\providecommand \bibitemNoStop [0]{.\EOS\space}%
\providecommand \EOS [0]{\spacefactor3000\relax}%
\providecommand \BibitemShut  [1]{\csname bibitem#1\endcsname}%
\let\auto@bib@innerbib\@empty
\bibitem [{\citenamefont {Bell}(1964)}]{bell64}%
  \BibitemOpen
  \bibfield  {author} {\bibinfo {author} {\bibfnamefont {J.~S.}\ \bibnamefont
  {Bell}},\ }\href@noop {} {\bibfield  {journal} {\bibinfo  {journal}
  {Physics}\ }\textbf {\bibinfo {volume} {1}},\ \bibinfo {pages} {195}
  (\bibinfo {year} {1964})}\BibitemShut {NoStop}%
\bibitem [{\citenamefont {Brunner}\ \emph {et~al.}(2014)\citenamefont
  {Brunner}, \citenamefont {Cavalcanti}, \citenamefont {Pironio}, \citenamefont
  {Scarani},\ and\ \citenamefont {Wehner}}]{BNL}%
  \BibitemOpen
  \bibfield  {author} {\bibinfo {author} {\bibfnamefont {N.}~\bibnamefont
  {Brunner}}, \bibinfo {author} {\bibfnamefont {D.}~\bibnamefont {Cavalcanti}},
  \bibinfo {author} {\bibfnamefont {S.}~\bibnamefont {Pironio}}, \bibinfo
  {author} {\bibfnamefont {V.}~\bibnamefont {Scarani}}, \ and\ \bibinfo
  {author} {\bibfnamefont {S.}~\bibnamefont {Wehner}},\ }\href {\doibase
  10.1103/RevModPhys.86.419} {\bibfield  {journal} {\bibinfo  {journal} {Rev.
  Mod. Phys.}\ }\textbf {\bibinfo {volume} {86}},\ \bibinfo {pages} {419}
  (\bibinfo {year} {2014})}\BibitemShut {NoStop}%
\bibitem [{\citenamefont {Svetlichny}(1987)}]{SI}%
  \BibitemOpen
  \bibfield  {author} {\bibinfo {author} {\bibfnamefont {G.}~\bibnamefont
  {Svetlichny}},\ }\href {\doibase 10.1103/PhysRevD.35.3066} {\bibfield
  {journal} {\bibinfo  {journal} {Phys. Rev. D}\ }\textbf {\bibinfo {volume}
  {35}},\ \bibinfo {pages} {3066} (\bibinfo {year} {1987})}\BibitemShut
  {NoStop}%
\bibitem [{\citenamefont {Popescu}\ and\ \citenamefont {Rohrlich}(1994)}]{PR}%
  \BibitemOpen
  \bibfield  {author} {\bibinfo {author} {\bibfnamefont {S.}~\bibnamefont
  {Popescu}}\ and\ \bibinfo {author} {\bibfnamefont {D.}~\bibnamefont
  {Rohrlich}},\ }\href {\doibase 10.1007/BF02058098} {\bibfield  {journal}
  {\bibinfo  {journal} {Found. Phys.}\ }\textbf {\bibinfo {volume} {24}},\
  \bibinfo {pages} {379} (\bibinfo {year} {1994})}\BibitemShut {NoStop}%
\bibitem [{\citenamefont {Masanes}\ \emph {et~al.}(2006)\citenamefont
  {Masanes}, \citenamefont {Acin},\ and\ \citenamefont {Gisin}}]{MAG06}%
  \BibitemOpen
  \bibfield  {author} {\bibinfo {author} {\bibfnamefont {L.}~\bibnamefont
  {Masanes}}, \bibinfo {author} {\bibfnamefont {A.}~\bibnamefont {Acin}}, \
  and\ \bibinfo {author} {\bibfnamefont {N.}~\bibnamefont {Gisin}},\ }\href
  {\doibase 10.1103/PhysRevA.73.012112} {\bibfield  {journal} {\bibinfo
  {journal} {Phys. Rev. A}\ }\textbf {\bibinfo {volume} {73}},\ \bibinfo
  {pages} {012112} (\bibinfo {year} {2006})}\BibitemShut {NoStop}%
\bibitem [{\citenamefont {Popescu}(2014)}]{RevSP}%
  \BibitemOpen
  \bibfield  {author} {\bibinfo {author} {\bibfnamefont {S.}~\bibnamefont
  {Popescu}},\ }\href {\doibase 10.1038/nphys2916} {\bibfield  {journal}
  {\bibinfo  {journal} {Nat. Phys.}\ }\textbf {\bibinfo {volume} {10}},\
  \bibinfo {pages} {264} (\bibinfo {year} {2014})}\BibitemShut {NoStop}%
\bibitem [{\citenamefont {Ollivier}\ and\ \citenamefont {Zurek}(2001)}]{WZQD}%
  \BibitemOpen
  \bibfield  {author} {\bibinfo {author} {\bibfnamefont {H.}~\bibnamefont
  {Ollivier}}\ and\ \bibinfo {author} {\bibfnamefont {W.~H.}\ \bibnamefont
  {Zurek}},\ }\href {\doibase 10.1103/PhysRevLett.88.017901} {\bibfield
  {journal} {\bibinfo  {journal} {Phys. Rev. Lett.}\ }\textbf {\bibinfo
  {volume} {88}},\ \bibinfo {pages} {017901} (\bibinfo {year}
  {2001})}\BibitemShut {NoStop}%
\bibitem [{\citenamefont {Henderson}\ and\ \citenamefont
  {Vedral}(2001)}]{WZQD1}%
  \BibitemOpen
  \bibfield  {author} {\bibinfo {author} {\bibfnamefont {L.}~\bibnamefont
  {Henderson}}\ and\ \bibinfo {author} {\bibfnamefont {V.}~\bibnamefont
  {Vedral}},\ }\href {http://stacks.iop.org/0305-4470/34/i=35/a=315} {\bibfield
   {journal} {\bibinfo  {journal} {J. Phys. A}\ }\textbf {\bibinfo {volume}
  {34}},\ \bibinfo {pages} {6899} (\bibinfo {year} {2001})}\BibitemShut
  {NoStop}%
\bibitem [{\citenamefont {Modi}\ \emph {et~al.}(2012)\citenamefont {Modi},
  \citenamefont {Brodutch}, \citenamefont {Cable}, \citenamefont {Paterek},\
  and\ \citenamefont {Vedral}}]{Modietal}%
  \BibitemOpen
  \bibfield  {author} {\bibinfo {author} {\bibfnamefont {K.}~\bibnamefont
  {Modi}}, \bibinfo {author} {\bibfnamefont {A.}~\bibnamefont {Brodutch}},
  \bibinfo {author} {\bibfnamefont {H.}~\bibnamefont {Cable}}, \bibinfo
  {author} {\bibfnamefont {T.}~\bibnamefont {Paterek}}, \ and\ \bibinfo
  {author} {\bibfnamefont {V.}~\bibnamefont {Vedral}},\ }\href {\doibase
  10.1103/RevModPhys.84.1655} {\bibfield  {journal} {\bibinfo  {journal} {Rev.
  Mod. Phys.}\ }\textbf {\bibinfo {volume} {84}},\ \bibinfo {pages} {1655}
  (\bibinfo {year} {2012})}\BibitemShut {NoStop}%
\bibitem [{\citenamefont {Cavalcanti}\ \emph {et~al.}(2011)\citenamefont
  {Cavalcanti}, \citenamefont {Aolita}, \citenamefont {Boixo}, \citenamefont
  {Modi}, \citenamefont {Piani},\ and\ \citenamefont {Winter}}]{QSM}%
  \BibitemOpen
  \bibfield  {author} {\bibinfo {author} {\bibfnamefont {D.}~\bibnamefont
  {Cavalcanti}}, \bibinfo {author} {\bibfnamefont {L.}~\bibnamefont {Aolita}},
  \bibinfo {author} {\bibfnamefont {S.}~\bibnamefont {Boixo}}, \bibinfo
  {author} {\bibfnamefont {K.}~\bibnamefont {Modi}}, \bibinfo {author}
  {\bibfnamefont {M.}~\bibnamefont {Piani}}, \ and\ \bibinfo {author}
  {\bibfnamefont {A.}~\bibnamefont {Winter}},\ }\href {\doibase
  10.1103/PhysRevA.83.032324} {\bibfield  {journal} {\bibinfo  {journal} {Phys.
  Rev. A}\ }\textbf {\bibinfo {volume} {83}},\ \bibinfo {pages} {032324}
  (\bibinfo {year} {2011})}\BibitemShut {NoStop}%
\bibitem [{\citenamefont {Dakic}\ \emph {et~al.}(2012)\citenamefont {Dakic},
  \citenamefont {Lipp}, \citenamefont {Ma}, \citenamefont {Ringbauer},
  \citenamefont {Kropatschek}, \citenamefont {Barz}, \citenamefont {Paterek},
  \citenamefont {Vedral}, \citenamefont {Zeilinger}, \citenamefont {Brukner},\
  and\ \citenamefont {Walther}}]{QDRS}%
  \BibitemOpen
  \bibfield  {author} {\bibinfo {author} {\bibfnamefont {B.}~\bibnamefont
  {Dakic}}, \bibinfo {author} {\bibfnamefont {Y.~O.}\ \bibnamefont {Lipp}},
  \bibinfo {author} {\bibfnamefont {X.}~\bibnamefont {Ma}}, \bibinfo {author}
  {\bibfnamefont {M.}~\bibnamefont {Ringbauer}}, \bibinfo {author}
  {\bibfnamefont {S.}~\bibnamefont {Kropatschek}}, \bibinfo {author}
  {\bibfnamefont {S.}~\bibnamefont {Barz}}, \bibinfo {author} {\bibfnamefont
  {T.}~\bibnamefont {Paterek}}, \bibinfo {author} {\bibfnamefont
  {V.}~\bibnamefont {Vedral}}, \bibinfo {author} {\bibfnamefont
  {A.}~\bibnamefont {Zeilinger}}, \bibinfo {author} {\bibfnamefont
  {C.}~\bibnamefont {Brukner}}, \ and\ \bibinfo {author} {\bibfnamefont
  {P.}~\bibnamefont {Walther}},\ }\href {\doibase 10.1038/nphys2377} {\bibfield
   {journal} {\bibinfo  {journal} {Nat. Phys.}\ }\textbf {\bibinfo {volume}
  {8}},\ \bibinfo {pages} {1745} (\bibinfo {year} {2012})}\BibitemShut
  {NoStop}%
\bibitem [{\citenamefont {Piani}\ \emph {et~al.}(2008)\citenamefont {Piani},
  \citenamefont {Horodecki},\ and\ \citenamefont {Horodecki}}]{CQstate}%
  \BibitemOpen
  \bibfield  {author} {\bibinfo {author} {\bibfnamefont {M.}~\bibnamefont
  {Piani}}, \bibinfo {author} {\bibfnamefont {P.}~\bibnamefont {Horodecki}}, \
  and\ \bibinfo {author} {\bibfnamefont {R.}~\bibnamefont {Horodecki}},\ }\href
  {\doibase 10.1103/PhysRevLett.100.090502} {\bibfield  {journal} {\bibinfo
  {journal} {Phys. Rev. Lett.}\ }\textbf {\bibinfo {volume} {100}},\ \bibinfo
  {pages} {090502} (\bibinfo {year} {2008})}\BibitemShut {NoStop}%
\bibitem [{\citenamefont {Daki\ifmmode~\acute{c}\else \'{c}\fi{}}\ \emph
  {et~al.}(2010)\citenamefont {Daki\ifmmode~\acute{c}\else \'{c}\fi{}},
  \citenamefont {Vedral},\ and\ \citenamefont {Brukner}}]{Dakicetal}%
  \BibitemOpen
  \bibfield  {author} {\bibinfo {author} {\bibfnamefont {B.}~\bibnamefont
  {Daki\ifmmode~\acute{c}\else \'{c}\fi{}}}, \bibinfo {author} {\bibfnamefont
  {V.}~\bibnamefont {Vedral}}, \ and\ \bibinfo {author} {\bibfnamefont
  {i.~c.~v.}\ \bibnamefont {Brukner}},\ }\href {\doibase
  10.1103/PhysRevLett.105.190502} {\bibfield  {journal} {\bibinfo  {journal}
  {Phys. Rev. Lett.}\ }\textbf {\bibinfo {volume} {105}},\ \bibinfo {pages}
  {190502} (\bibinfo {year} {2010})}\BibitemShut {NoStop}%
\bibitem [{\citenamefont {Giorgi}\ \emph {et~al.}(2011)\citenamefont {Giorgi},
  \citenamefont {Bellomo}, \citenamefont {Galve},\ and\ \citenamefont
  {Zambrini}}]{GTC}%
  \BibitemOpen
  \bibfield  {author} {\bibinfo {author} {\bibfnamefont {G.~L.}\ \bibnamefont
  {Giorgi}}, \bibinfo {author} {\bibfnamefont {B.}~\bibnamefont {Bellomo}},
  \bibinfo {author} {\bibfnamefont {F.}~\bibnamefont {Galve}}, \ and\ \bibinfo
  {author} {\bibfnamefont {R.}~\bibnamefont {Zambrini}},\ }\href {\doibase
  10.1103/PhysRevLett.107.190501} {\bibfield  {journal} {\bibinfo  {journal}
  {Phys. Rev. Lett.}\ }\textbf {\bibinfo {volume} {107}},\ \bibinfo {pages}
  {190501} (\bibinfo {year} {2011})}\BibitemShut {NoStop}%
\bibitem [{\citenamefont {Zhao}\ \emph {et~al.}(2013)\citenamefont {Zhao},
  \citenamefont {Hu}, \citenamefont {Yue},\ and\ \citenamefont {Fan}}]{GCTS}%
  \BibitemOpen
  \bibfield  {author} {\bibinfo {author} {\bibfnamefont {L.}~\bibnamefont
  {Zhao}}, \bibinfo {author} {\bibfnamefont {X.}~\bibnamefont {Hu}}, \bibinfo
  {author} {\bibfnamefont {R.-H.}\ \bibnamefont {Yue}}, \ and\ \bibinfo
  {author} {\bibfnamefont {H.}~\bibnamefont {Fan}},\ }\href {\doibase
  10.1007/s11128-013-0525-9} {\bibfield  {journal} {\bibinfo  {journal}
  {Quantum Inf. Process}\ }\textbf {\bibinfo {volume} {12}},\ \bibinfo {pages}
  {2371} (\bibinfo {year} {2013})}\BibitemShut {NoStop}%
\bibitem [{\citenamefont {Beggi}\ \emph {et~al.}(2015)\citenamefont {Beggi},
  \citenamefont {Buscemi},\ and\ \citenamefont {Bordone}}]{GTQD}%
  \BibitemOpen
  \bibfield  {author} {\bibinfo {author} {\bibfnamefont {A.}~\bibnamefont
  {Beggi}}, \bibinfo {author} {\bibfnamefont {F.}~\bibnamefont {Buscemi}}, \
  and\ \bibinfo {author} {\bibfnamefont {P.}~\bibnamefont {Bordone}},\ }\href
  {\doibase 10.1007/s11128-014-0882-z} {\bibfield  {journal} {\bibinfo
  {journal} {Quantum Inf. Process}\ }\textbf {\bibinfo {volume} {14}},\
  \bibinfo {pages} {573} (\bibinfo {year} {2015})}\BibitemShut {NoStop}%
\bibitem [{\citenamefont {Barrett}\ \emph {et~al.}(2005)\citenamefont
  {Barrett}, \citenamefont {Linden}, \citenamefont {Massar}, \citenamefont
  {Pironio}, \citenamefont {Popescu},\ and\ \citenamefont {Roberts}}]{Barrett}%
  \BibitemOpen
  \bibfield  {author} {\bibinfo {author} {\bibfnamefont {J.}~\bibnamefont
  {Barrett}}, \bibinfo {author} {\bibfnamefont {N.}~\bibnamefont {Linden}},
  \bibinfo {author} {\bibfnamefont {S.}~\bibnamefont {Massar}}, \bibinfo
  {author} {\bibfnamefont {S.}~\bibnamefont {Pironio}}, \bibinfo {author}
  {\bibfnamefont {S.}~\bibnamefont {Popescu}}, \ and\ \bibinfo {author}
  {\bibfnamefont {D.}~\bibnamefont {Roberts}},\ }\href {\doibase
  10.1103/PhysRevA.71.022101} {\bibfield  {journal} {\bibinfo  {journal} {Phys.
  Rev. A}\ }\textbf {\bibinfo {volume} {71}},\ \bibinfo {pages} {022101}
  (\bibinfo {year} {2005})}\BibitemShut {NoStop}%
\bibitem [{\citenamefont {Mermin}(1990)}]{mermin}%
  \BibitemOpen
  \bibfield  {author} {\bibinfo {author} {\bibfnamefont {N.~D.}\ \bibnamefont
  {Mermin}},\ }\href {\doibase 10.1103/PhysRevLett.65.1838} {\bibfield
  {journal} {\bibinfo  {journal} {Phys. Rev. Lett.}\ }\textbf {\bibinfo
  {volume} {65}},\ \bibinfo {pages} {1838} (\bibinfo {year}
  {1990})}\BibitemShut {NoStop}%
\bibitem [{\citenamefont {Seevinck}\ and\ \citenamefont {Uffink}(2001)}]{SU01}%
  \BibitemOpen
  \bibfield  {author} {\bibinfo {author} {\bibfnamefont {M.}~\bibnamefont
  {Seevinck}}\ and\ \bibinfo {author} {\bibfnamefont {J.}~\bibnamefont
  {Uffink}},\ }\href {\doibase 10.1103/PhysRevA.65.012107} {\bibfield
  {journal} {\bibinfo  {journal} {Phys. Rev. A}\ }\textbf {\bibinfo {volume}
  {65}},\ \bibinfo {pages} {012107} (\bibinfo {year} {2001})}\BibitemShut
  {NoStop}%
\bibitem [{\citenamefont {Collins}\ \emph {et~al.}(2002)\citenamefont
  {Collins}, \citenamefont {Gisin}, \citenamefont {Popescu}, \citenamefont
  {Roberts},\ and\ \citenamefont {Scarani}}]{CGP+02}%
  \BibitemOpen
  \bibfield  {author} {\bibinfo {author} {\bibfnamefont {D.}~\bibnamefont
  {Collins}}, \bibinfo {author} {\bibfnamefont {N.}~\bibnamefont {Gisin}},
  \bibinfo {author} {\bibfnamefont {S.}~\bibnamefont {Popescu}}, \bibinfo
  {author} {\bibfnamefont {D.}~\bibnamefont {Roberts}}, \ and\ \bibinfo
  {author} {\bibfnamefont {V.}~\bibnamefont {Scarani}},\ }\href {\doibase
  10.1103/PhysRevLett.88.170405} {\bibfield  {journal} {\bibinfo  {journal}
  {Phys. Rev. Lett.}\ }\textbf {\bibinfo {volume} {88}},\ \bibinfo {pages}
  {170405} (\bibinfo {year} {2002})}\BibitemShut {NoStop}%
\bibitem [{\citenamefont {Bancal}\ \emph {et~al.}(2011)\citenamefont {Bancal},
  \citenamefont {Gisin}, \citenamefont {Liang},\ and\ \citenamefont
  {Pironio}}]{DIMulti}%
  \BibitemOpen
  \bibfield  {author} {\bibinfo {author} {\bibfnamefont {J.-D.}\ \bibnamefont
  {Bancal}}, \bibinfo {author} {\bibfnamefont {N.}~\bibnamefont {Gisin}},
  \bibinfo {author} {\bibfnamefont {Y.-C.}\ \bibnamefont {Liang}}, \ and\
  \bibinfo {author} {\bibfnamefont {S.}~\bibnamefont {Pironio}},\ }\href
  {\doibase 10.1103/PhysRevLett.106.250404} {\bibfield  {journal} {\bibinfo
  {journal} {Phys. Rev. Lett.}\ }\textbf {\bibinfo {volume} {106}},\ \bibinfo
  {pages} {250404} (\bibinfo {year} {2011})}\BibitemShut {NoStop}%
\bibitem [{\citenamefont {Donohue}\ and\ \citenamefont {Wolfe}(2015)}]{suplo}%
  \BibitemOpen
  \bibfield  {author} {\bibinfo {author} {\bibfnamefont {J.~M.}\ \bibnamefont
  {Donohue}}\ and\ \bibinfo {author} {\bibfnamefont {E.}~\bibnamefont
  {Wolfe}},\ }\href {\doibase 10.1103/PhysRevA.92.062120} {\bibfield  {journal}
  {\bibinfo  {journal} {Phys. Rev. A}\ }\textbf {\bibinfo {volume} {92}},\
  \bibinfo {pages} {062120} (\bibinfo {year} {2015})}\BibitemShut {NoStop}%
\bibitem [{\citenamefont {Elitzur}\ \emph {et~al.}(1992)\citenamefont
  {Elitzur}, \citenamefont {Popescu},\ and\ \citenamefont {Rohrlich}}]{EPR2}%
  \BibitemOpen
  \bibfield  {author} {\bibinfo {author} {\bibfnamefont {A.~C.}\ \bibnamefont
  {Elitzur}}, \bibinfo {author} {\bibfnamefont {S.}~\bibnamefont {Popescu}}, \
  and\ \bibinfo {author} {\bibfnamefont {D.}~\bibnamefont {Rohrlich}},\ }\href
  {\doibase http://dx.doi.org/10.1016/0375-9601(92)90952-I} {\bibfield
  {journal} {\bibinfo  {journal} {Phys. Lett. A}\ }\textbf {\bibinfo {volume}
  {162}},\ \bibinfo {pages} {25 } (\bibinfo {year} {1992})}\BibitemShut
  {NoStop}%
\bibitem [{\citenamefont {Fine}(1982)}]{LHV}%
  \BibitemOpen
  \bibfield  {author} {\bibinfo {author} {\bibfnamefont {A.}~\bibnamefont
  {Fine}},\ }\href {\doibase 10.1103/PhysRevLett.48.291} {\bibfield  {journal}
  {\bibinfo  {journal} {Phys. Rev. Lett.}\ }\textbf {\bibinfo {volume} {48}},\
  \bibinfo {pages} {291} (\bibinfo {year} {1982})}\BibitemShut {NoStop}%
\bibitem [{\citenamefont {Werner}\ and\ \citenamefont
  {Wolf}(2001{\natexlab{a}})}]{WernerWolfmulti}%
  \BibitemOpen
  \bibfield  {author} {\bibinfo {author} {\bibfnamefont {R.~F.}\ \bibnamefont
  {Werner}}\ and\ \bibinfo {author} {\bibfnamefont {M.~M.}\ \bibnamefont
  {Wolf}},\ }\href {\doibase 10.1103/PhysRevA.64.032112} {\bibfield  {journal}
  {\bibinfo  {journal} {Phys. Rev. A}\ }\textbf {\bibinfo {volume} {64}},\
  \bibinfo {pages} {032112} (\bibinfo {year} {2001}{\natexlab{a}})}\BibitemShut
  {NoStop}%
\bibitem [{\citenamefont {Pironio}\ \emph {et~al.}(2011)\citenamefont
  {Pironio}, \citenamefont {Bancal},\ and\ \citenamefont
  {Scarani}}]{Pironioetal}%
  \BibitemOpen
  \bibfield  {author} {\bibinfo {author} {\bibfnamefont {S.}~\bibnamefont
  {Pironio}}, \bibinfo {author} {\bibfnamefont {J.-D.}\ \bibnamefont {Bancal}},
  \ and\ \bibinfo {author} {\bibfnamefont {V.}~\bibnamefont {Scarani}},\ }\href
  {http://stacks.iop.org/1751-8121/44/i=6/a=065303} {\bibfield  {journal}
  {\bibinfo  {journal} {J. Phys. A}\ }\textbf {\bibinfo {volume} {44}},\
  \bibinfo {pages} {065303} (\bibinfo {year} {2011})}\BibitemShut {NoStop}%
\bibitem [{\citenamefont {Bancal}\ \emph {et~al.}(2013)\citenamefont {Bancal},
  \citenamefont {Barrett}, \citenamefont {Gisin},\ and\ \citenamefont
  {Pironio}}]{Banceletal}%
  \BibitemOpen
  \bibfield  {author} {\bibinfo {author} {\bibfnamefont {J.-D.}\ \bibnamefont
  {Bancal}}, \bibinfo {author} {\bibfnamefont {J.}~\bibnamefont {Barrett}},
  \bibinfo {author} {\bibfnamefont {N.}~\bibnamefont {Gisin}}, \ and\ \bibinfo
  {author} {\bibfnamefont {S.}~\bibnamefont {Pironio}},\ }\href {\doibase
  10.1103/PhysRevA.88.014102} {\bibfield  {journal} {\bibinfo  {journal} {Phys.
  Rev. A}\ }\textbf {\bibinfo {volume} {88}},\ \bibinfo {pages} {014102}
  (\bibinfo {year} {2013})}\BibitemShut {NoStop}%
\bibitem [{\citenamefont {Ac\'{\i}n}\ \emph {et~al.}(2006)\citenamefont
  {Ac\'{\i}n}, \citenamefont {Gisin},\ and\ \citenamefont {Masanes}}]{DQKD}%
  \BibitemOpen
  \bibfield  {author} {\bibinfo {author} {\bibfnamefont {A.}~\bibnamefont
  {Ac\'{\i}n}}, \bibinfo {author} {\bibfnamefont {N.}~\bibnamefont {Gisin}}, \
  and\ \bibinfo {author} {\bibfnamefont {L.}~\bibnamefont {Masanes}},\ }\href
  {\doibase 10.1103/PhysRevLett.97.120405} {\bibfield  {journal} {\bibinfo
  {journal} {Phys. Rev. Lett.}\ }\textbf {\bibinfo {volume} {97}},\ \bibinfo
  {pages} {120405} (\bibinfo {year} {2006})}\BibitemShut {NoStop}%
\bibitem [{\citenamefont {Clauser}\ \emph {et~al.}(1969)\citenamefont
  {Clauser}, \citenamefont {Horne}, \citenamefont {Shimony},\ and\
  \citenamefont {Holt}}]{chsh}%
  \BibitemOpen
  \bibfield  {author} {\bibinfo {author} {\bibfnamefont {J.~F.}\ \bibnamefont
  {Clauser}}, \bibinfo {author} {\bibfnamefont {M.~A.}\ \bibnamefont {Horne}},
  \bibinfo {author} {\bibfnamefont {A.}~\bibnamefont {Shimony}}, \ and\
  \bibinfo {author} {\bibfnamefont {R.~A.}\ \bibnamefont {Holt}},\ }\href
  {\doibase 10.1103/PhysRevLett.23.880} {\bibfield  {journal} {\bibinfo
  {journal} {Phys. Rev. Lett.}\ }\textbf {\bibinfo {volume} {23}},\ \bibinfo
  {pages} {880} (\bibinfo {year} {1969})}\BibitemShut {NoStop}%
\bibitem [{\citenamefont {Werner}\ and\ \citenamefont
  {Wolf}(2001{\natexlab{b}})}]{WernerWolf}%
  \BibitemOpen
  \bibfield  {author} {\bibinfo {author} {\bibfnamefont {R.~F.}\ \bibnamefont
  {Werner}}\ and\ \bibinfo {author} {\bibfnamefont {M.~M.}\ \bibnamefont
  {Wolf}},\ }\href@noop {} {\bibfield  {journal} {\bibinfo  {journal} {Quantum
  Inf. Comput.}\ }\textbf {\bibinfo {volume} {1}},\ \bibinfo {pages} {1}
  (\bibinfo {year} {2001}{\natexlab{b}})}\BibitemShut {NoStop}%
\bibitem [{\citenamefont {Brunner}\ \emph {et~al.}(2011)\citenamefont
  {Brunner}, \citenamefont {Cavalcanti}, \citenamefont {Salles},\ and\
  \citenamefont {Skrzypczyk}}]{EPR2B}%
  \BibitemOpen
  \bibfield  {author} {\bibinfo {author} {\bibfnamefont {N.}~\bibnamefont
  {Brunner}}, \bibinfo {author} {\bibfnamefont {D.}~\bibnamefont {Cavalcanti}},
  \bibinfo {author} {\bibfnamefont {A.}~\bibnamefont {Salles}}, \ and\ \bibinfo
  {author} {\bibfnamefont {P.}~\bibnamefont {Skrzypczyk}},\ }\href {\doibase
  10.1103/PhysRevLett.106.020402} {\bibfield  {journal} {\bibinfo  {journal}
  {Phys. Rev. Lett.}\ }\textbf {\bibinfo {volume} {106}},\ \bibinfo {pages}
  {020402} (\bibinfo {year} {2011})}\BibitemShut {NoStop}%
\bibitem [{\citenamefont {Greenberger}\ \emph {et~al.}(1989)\citenamefont
  {Greenberger}, \citenamefont {Horne},\ and\ \citenamefont {Zeilinger}}]{GHZ}%
  \BibitemOpen
  \bibfield  {author} {\bibinfo {author} {\bibfnamefont {D.~M.}\ \bibnamefont
  {Greenberger}}, \bibinfo {author} {\bibfnamefont {M.~A.}\ \bibnamefont
  {Horne}}, \ and\ \bibinfo {author} {\bibfnamefont {A.}~\bibnamefont
  {Zeilinger}},\ }\href@noop {} {\emph {\bibinfo {title} {in Bell's Theorem,
  Quantum Theory, and Conceptions of the Universe}}},\ edited by\ \bibinfo
  {editor} {\bibfnamefont {M.}~\bibnamefont {Kafatos}}\ (\bibinfo  {publisher}
  {Kluwer Academic Dordrecht},\ \bibinfo {year} {1989})\BibitemShut {NoStop}%
\bibitem [{\citenamefont {Coffman}\ \emph {et~al.}(2000)\citenamefont
  {Coffman}, \citenamefont {Kundu},\ and\ \citenamefont {Wootters}}]{CKW}%
  \BibitemOpen
  \bibfield  {author} {\bibinfo {author} {\bibfnamefont {V.}~\bibnamefont
  {Coffman}}, \bibinfo {author} {\bibfnamefont {J.}~\bibnamefont {Kundu}}, \
  and\ \bibinfo {author} {\bibfnamefont {W.~K.}\ \bibnamefont {Wootters}},\
  }\href {\doibase 10.1103/PhysRevA.61.052306} {\bibfield  {journal} {\bibinfo
  {journal} {Phys. Rev. A}\ }\textbf {\bibinfo {volume} {61}},\ \bibinfo
  {pages} {052306} (\bibinfo {year} {2000})}\BibitemShut {NoStop}%
\bibitem [{\citenamefont {Wootters}(1998)}]{WKW}%
  \BibitemOpen
  \bibfield  {author} {\bibinfo {author} {\bibfnamefont {W.~K.}\ \bibnamefont
  {Wootters}},\ }\href {\doibase 10.1103/PhysRevLett.80.2245} {\bibfield
  {journal} {\bibinfo  {journal} {Phys. Rev. Lett.}\ }\textbf {\bibinfo
  {volume} {80}},\ \bibinfo {pages} {2245} (\bibinfo {year}
  {1998})}\BibitemShut {NoStop}%
\bibitem [{\citenamefont {Jebaratnam}(2016{\natexlab{a}})}]{JebaEPR}%
  \BibitemOpen
  \bibfield  {author} {\bibinfo {author} {\bibfnamefont {C.}~\bibnamefont
  {Jebaratnam}},\ }\href {\doibase 10.1103/PhysRevA.93.052311} {\bibfield
  {journal} {\bibinfo  {journal} {Phys. Rev. A}\ }\textbf {\bibinfo {volume}
  {93}},\ \bibinfo {pages} {052311} (\bibinfo {year}
  {2016}{\natexlab{a}})}\BibitemShut {NoStop}%
\bibitem [{\citenamefont {Chi}\ \emph {et~al.}(2010)\citenamefont {Chi},
  \citenamefont {Jeong}, \citenamefont {Kim}, \citenamefont {Lee},\ and\
  \citenamefont {Lee}}]{Chietal}%
  \BibitemOpen
  \bibfield  {author} {\bibinfo {author} {\bibfnamefont {D.~P.}\ \bibnamefont
  {Chi}}, \bibinfo {author} {\bibfnamefont {K.}~\bibnamefont {Jeong}}, \bibinfo
  {author} {\bibfnamefont {T.}~\bibnamefont {Kim}}, \bibinfo {author}
  {\bibfnamefont {K.}~\bibnamefont {Lee}}, \ and\ \bibinfo {author}
  {\bibfnamefont {S.}~\bibnamefont {Lee}},\ }\href {\doibase
  10.1103/PhysRevA.81.044302} {\bibfield  {journal} {\bibinfo  {journal} {Phys.
  Rev. A}\ }\textbf {\bibinfo {volume} {81}},\ \bibinfo {pages} {044302}
  (\bibinfo {year} {2010})}\BibitemShut {NoStop}%
\bibitem [{\citenamefont {Ajoy}\ and\ \citenamefont {Rungta}(2010)}]{Ajoy}%
  \BibitemOpen
  \bibfield  {author} {\bibinfo {author} {\bibfnamefont {A.}~\bibnamefont
  {Ajoy}}\ and\ \bibinfo {author} {\bibfnamefont {P.}~\bibnamefont {Rungta}},\
  }\href {\doibase 10.1103/PhysRevA.81.052334} {\bibfield  {journal} {\bibinfo
  {journal} {Phys. Rev. A}\ }\textbf {\bibinfo {volume} {81}},\ \bibinfo
  {pages} {052334} (\bibinfo {year} {2010})}\BibitemShut {NoStop}%
\bibitem [{\citenamefont {Horodecki}\ \emph {et~al.}(2009)\citenamefont
  {Horodecki}, \citenamefont {Horodecki}, \citenamefont {Horodecki},\ and\
  \citenamefont {Horodecki}}]{Horodecki}%
  \BibitemOpen
  \bibfield  {author} {\bibinfo {author} {\bibfnamefont {R.}~\bibnamefont
  {Horodecki}}, \bibinfo {author} {\bibfnamefont {P.}~\bibnamefont
  {Horodecki}}, \bibinfo {author} {\bibfnamefont {M.}~\bibnamefont
  {Horodecki}}, \ and\ \bibinfo {author} {\bibfnamefont {K.}~\bibnamefont
  {Horodecki}},\ }\href {\doibase 10.1103/RevModPhys.81.865} {\bibfield
  {journal} {\bibinfo  {journal} {Rev. Mod. Phys.}\ }\textbf {\bibinfo {volume}
  {81}},\ \bibinfo {pages} {865} (\bibinfo {year} {2009})}\BibitemShut
  {NoStop}%
\bibitem [{\citenamefont {Bennett}\ \emph {et~al.}(2011)\citenamefont
  {Bennett}, \citenamefont {Grudka}, \citenamefont {Horodecki}, \citenamefont
  {Horodecki},\ and\ \citenamefont {Horodecki}}]{FCBox}%
  \BibitemOpen
  \bibfield  {author} {\bibinfo {author} {\bibfnamefont {C.~H.}\ \bibnamefont
  {Bennett}}, \bibinfo {author} {\bibfnamefont {A.}~\bibnamefont {Grudka}},
  \bibinfo {author} {\bibfnamefont {M.}~\bibnamefont {Horodecki}}, \bibinfo
  {author} {\bibfnamefont {P.}~\bibnamefont {Horodecki}}, \ and\ \bibinfo
  {author} {\bibfnamefont {R.}~\bibnamefont {Horodecki}},\ }\href {\doibase
  10.1103/PhysRevA.83.012312} {\bibfield  {journal} {\bibinfo  {journal} {Phys.
  Rev. A}\ }\textbf {\bibinfo {volume} {83}},\ \bibinfo {pages} {012312}
  (\bibinfo {year} {2011})}\BibitemShut {NoStop}%
\bibitem [{\citenamefont {Jebaratnam}()}]{Jeba1}%
  \BibitemOpen
  \bibfield  {author} {\bibinfo {author} {\bibfnamefont {C.}~\bibnamefont
  {Jebaratnam}},\ }\href {https://arxiv.org/abs/1407.3170} {\enquote {\bibinfo
  {title} {Canonical decomposition of quantum correlations in the framework of
  generalized nonsignaling theories},}\ }\bibinfo {note}
  {ArXiv:1407.3170v4}\BibitemShut {NoStop}%
\bibitem [{\citenamefont {Jebaratnam}(2016{\natexlab{b}})}]{Jeba}%
  \BibitemOpen
  \bibfield  {author} {\bibinfo {author} {\bibfnamefont {C.}~\bibnamefont
  {Jebaratnam}},\ }\emph {\bibinfo {title} {Characterizing quantum correlations
  in the nonsignaling framework}},\ \href {https://arxiv.org/abs/1605.06445}
  {Ph.D. thesis},\ \bibinfo  {school} {Indian Institute of Science Education
  and Research Mohali} (\bibinfo {year} {2016}{\natexlab{b}}),\ \bibinfo {note}
  {arXiv:1605.06445}\BibitemShut {NoStop}%
\bibitem [{\citenamefont {Mukherjee}\ \emph {et~al.}(2015)\citenamefont
  {Mukherjee}, \citenamefont {Paul},\ and\ \citenamefont {Sarkar}}]{gnl99}%
  \BibitemOpen
  \bibfield  {author} {\bibinfo {author} {\bibfnamefont {K.}~\bibnamefont
  {Mukherjee}}, \bibinfo {author} {\bibfnamefont {B.}~\bibnamefont {Paul}}, \
  and\ \bibinfo {author} {\bibfnamefont {D.}~\bibnamefont {Sarkar}},\ }\href
  {http://stacks.iop.org/1751-8121/48/i=46/a=465302} {\bibfield  {journal}
  {\bibinfo  {journal} {J. Phys. A}\ }\textbf {\bibinfo {volume} {48}},\
  \bibinfo {pages} {465302} (\bibinfo {year} {2015})}\BibitemShut {NoStop}%
\end{thebibliography}%
\end{document}